\documentclass[letterpaper,english]{article}
\usepackage[T1]{fontenc}
\usepackage[latin9]{inputenc}
\usepackage{float}
\usepackage{mathrsfs}
\usepackage{bm}
\usepackage{amsmath}
\usepackage{amssymb}
\usepackage{graphicx}
\usepackage{setspace}
\usepackage{dcolumn}
\newcolumntype{d}[1]{D{.}{.}{#1}}
\newcommand\mc[1]{\multicolumn{1}{c}{#1}}

\usepackage{url}

\makeatletter

\pdfpageheight\paperheight
\pdfpagewidth\paperwidth

\providecommand{\tabularnewline}{\\}
\floatstyle{ruled}
\newfloat{algorithm}{tbp}{loa}
\providecommand{\algorithmname}{Algorithm}
\floatname{algorithm}{\protect\algorithmname}

\date{}
\usepackage[margin=1in]{geometry}
\usepackage{amssymb}
\usepackage{babel}

\usepackage{algorithm,algpseudocode}

\usepackage{xfrac}

\makeatother

\usepackage{babel}

\begin{document}
\title{Near-Linear Orbit Uncertainty Propagation in the Perturbed Two-Body
Problem}
\author{Javier Hernando-Ayuso\thanks{Mission Analysis and Flight Dynamics
Engineer (in his personal capacity)},
Claudio Bombardelli\thanks{Associate Professor, Space Dynamics Group},
Giulio Ba\`u\thanks{Assistant Professor, University of Pisa}, Alicia
Mart\'inez-Cacho\thanks{PhD student, Space Dynamics Group}}
\maketitle
\begin{center}
\textsl{$^{*}$}\textit{ispace-inc, Tokyo, 113-0007,}{} Japan 
\par\end{center}

\begin{center}
\textsl{$^{\dagger}$$^{\mathsection}$}\textit{Technical University
of Madrid (UPM), Madrid, 28040,}{} Spain 
\par\end{center}

\begin{center}
\textsl{$^{\ddagger}$}\textit{University of Pisa, Pisa, 56126,}{}
Italy 
\par\end{center}

\begin{center}
\textbf{Abstract} 
\par\end{center}

The paper addresses the problem of minimizing the impact of
non-linearities when dealing with uncertainty propagation in the
perturbed two-body problem. The recently introduced generalized
equinoctial orbital element set (GEqOE) is employed as a means to
reduce non-linear effects stemming from J$_2$ and higher order gravity
field harmonics. The uncertainty propagation performance of the
proposed set of elements in different Earth orbit scenarios, including
low-thrust orbit raising, is evaluated using a Cram\'er-von Mises test
on the Mahalanobis distance of the uncertainty distribution. A
considerable improvement compared to all sets of elements proposed so
far is obtained.

\section{Introduction}

The ability to rapidly and accurately propagating the probability
density function (pdf) of a space object in time is key in space
situational awareness (SSA). Tracking an Earth orbiting object through
subsequent observation arcs or correctly estimating the probability of
collision between two objects sufficiently early in time, for
instance, are operations that require an accurate knowledge of the
temporal evolution of a pdf.

The propagation of the pdf that characterizes the uncertainty of the
orbital state vector is governed by the Fokker-Plank Equation
\cite[pp. 192-202]{Maybeck1982vol2}. This partial differential
equation is computationally hard to solve, which has motivated the use
of approximate methods that can compute the pdf evolution with
reasonable accuracy and computational cost.  A spectrum of
possibilities are available in the literature ranging from very fast
but less accurate linear methods to the highly accurate yet
computationally expensive Monte-Carlo-based methods. These include
Differential Algebra \cite{Armellin2010}, state transition tensors
\cite{Park2006,roa2021reduced}, Gaussian Mixture Models
\cite{Giza2009,terejanu2008uncertainty}, unscented transform
\cite{Julier1997}, Polynomial Chaos Expansion \cite{Jones2013}, Line
of Variations \cite{Milani2005}, Kriging
\cite{Tardioli2015}. Combinations of different approaches have also
been explored \cite{vittaldev2016spacecraft}. The common goal of these
techniques is to handle the effect of nonlinearities associated with
the perturbed orbital motion and resulting in a pdf evolution that can
rapidly become far from Gaussian.

Irrespectively of the strategy adopted to tackle nonlinear effects,
the choice of the mathematical formulation is a key element when
constructing an efficient uncertainty propagation method
(\cite{junkins1996non,Sabol2010,Folcik2011,HernandoAyuso2017,
  Coppola2015,Melton2000,Lane2006,Hill2008,Weis2018,Roa2015OD}). Arguably,
before even considering sophisticated algorithms like
\cite{Armellin2010}--\cite{vittaldev2016spacecraft} one should
investigate how to minimize non-linearities that are
\textit{intrinsic} to the mathematical structure of the propagated set
of ordinary differential equations. A significant step forward in this
direction has been made by Arisfoff et al. \cite{AHA_2021}, who
proposed a set of ``J$_2$ equinoctial orbital elements'' (J2EqOE)
based on Brouwer-Lyddane's solution of the main satellite problem. The
method allows one to absorb the non-linearities stemming from the
J$_2$ term of the geopotential perturbation and provides a major
improvement in uncertainty realism (UR) compared, for example, to
classical equinoctial elements.

The purpose of this article is to test the uncertainty propagation
performance of yet another set of elements, the generalized
equinoctial orbital elements (GEqOE), recently introduced by Ba\`u et
al. \cite{bau2021generalization}.  Similarly to Aristoff's elements,
GEqOE mitigate the negative effect of J$_2$ in the propagation of the
orbital state. However, there are important differences between the
two sets of elements. The first is that GEqOE are related to Cartesian
coordinates by transformations expressed in closed analytical form
(Sects. 3, 4 in \cite{bau2021generalization}) and evolve according to
a set of clearly defined differential equations of motion (Sect. 5 in
\cite{bau2021generalization}). The second is that they can be
constructed accounting for any perturbation deriving from a potential,
not just J$_2$. These characteristics make the new set of elements
particularly appealing for uncertainty propagation in low-Earth orbit
and motivate a detailed analysis of their performance compared to the
one of the already proposed J2EqOE.

The article is organized as follows. First a brief review of the
generalized orbital motion quantities and the corresponding orbital
elements introduced by Ba\`u et al. \cite{bau2021generalization} is
provided for convenience.  Next, the mathematical treatment of the
linear uncertainty propagation in GEqOE is developed. A fully
analytical explanation of the reduction, by use of the proposed
elements, of uncertainty propagation nonlinearities associated to
secular effects is provided in the subsequent section.  Finally, an
extensive simulation campaign is conducted to evaluate the UR of the
proposed elements compared to competing sets of elements proposed in
the literature. The test cases, which include orbits of different
eccentricities and inclinations and the impact of low-thrust
propulsion, are simulated with a high-fidelity model including
high-order geopotential harmonics and third-body effects.

\section{Generalized Orbital Motion Quantities}

Let us consider the perturbed two-body problem written in a geocentric
inertial reference frame:
\begin{equation}
  \ddot{\bf r}=-\frac{\mu\mathbf{r}}{r^{3}}+\mathbf{F}(\mathbf{r},\dot{\mathbf{r}},t),
  \label{eq:eqm}
\end{equation}
where ${\bf r}$, $\dot{\bf r}$, and $\ddot{\bf r}$ are the geocentric
position, velocity, and acceleration, respectively. Moreover, $r$ is
the position magnitude, $t$ denotes time, and $\mu$ is the
gravitational parameter of the Earth.

Following \cite{bau2021generalization}, $\mathbf{F}$ is split into a
term that is derivable from a potential energy
$\mathscr{U}(\mathbf{r},\,t)$ and a term
$\mathbf{P}(\mathbf{r},\dot{\mathbf{r}},t)$ that is not:
\begin{equation}
  \mathbf{F}=\mathbf{P}-\nabla\mathscr{U}.
  \label{eq:FP}
\end{equation}
Next, the \textit{total orbital energy} $\mathscr{E}$ is introduced by
adding to the Keplerian energy $\mathscr{E}_{K}$ the potential energy
$\mathscr{U}$:
\[
\mathscr{E}=\mathscr{E}_{K}+\mathscr{U}=\frac{v^{2}}{2}-\frac{\mu}{r}+\mathscr{U},
\]
where $v$ is the velocity magnitude.

By formally replacing the Keplerian energy with the total energy in
the momentum-energy relation as discussed in
\cite{bau2021generalization} one obtains the \textit{generalized
  angular momentum}
\begin{equation}
  c=r\sqrt{2\left(\mathscr{E}+\frac{\mu}{r}\right)-u^{2}}=\sqrt{h^{2}+2\mathscr{U}r^{2}},
  \label{eq:am_gen}
\end{equation}
where $u$ and $h$ are, respectively, the radial velocity and the
osculating angular momentum. This new quantity can be employed to
define the \textit{generalized angular momentum vector} and
\textit{eccentricity vector} as, respectively:
\begin{align*}
  {\bf c} & =c\,{\bf e}_{h},\\
  {\bf g} & =\frac{1}{\mu}{\bm{\upsilon}}\times{\bf c}-{\bf e}_{r},
\end{align*}
where 
\[
{\bm{\upsilon}}=u\,{\bf e}_{r}+\frac{c}{r}{\bf e}_{f},
\]
is the generalized velocity vector and $\left\{ {\bf e}_{r},\,{\bf
  e}_{f},\,{\bf e}_{h}\right\} $ is the \textit{orbital} reference
frame orthonormal basis.

Denoting by $g$ the generalized eccentricity, that is $g=\left|{\bf
  g}\right|$, it is found that $g$, $c$, $\mathscr{E}$ satisfy the
relation
\begin{equation}
  g=\frac{1}{\mu}\sqrt{\mu^{2}+2\mathscr{E}c^{2}}=\sqrt{e^{2}+\frac{2\mathscr{U}}{\mu^{2}}
    \left(h^{2}+2\mathscr{E}r^{2}\right)},
  \label{eq:ecc_gen}
\end{equation}
where $e$ is the osculating eccentricity. The vectors ${\bf c}$, ${\bf
  g}$ define at any time (as long as $c\ne0$) a \textit{non-osculating
  ellipse} $\Gamma$, which lies on the orbital plane and has one focus
located at the center of mass of the primary body of attraction
\cite{bau2021generalization}.  The semi-major axis of that conic is
the \textit{generalized semi-major axis} ($\textrm{a}$) and is related
to the osculating semi-major axis ($a$) by the formula
\begin{equation}
  \textsl{\textrm{a}}=-\frac{\mu}{2\mathscr{E}}=a+\frac{\mathscr{\mu U}}{2\mathscr{E}\mathscr{E}_{K}}.
  \label{eq:sa_gen}
\end{equation}
From equations \eqref{eq:am_gen}, \eqref{eq:ecc_gen},
\eqref{eq:sa_gen} it is immediate to see that $c$, $g$,
$\textsl{\textrm{a}}$ coincide with their osculating counterparts $h$,
$e$, $a$ when $\mathscr{U}=0$.

Assuming, from now on, that $\mathscr{E}<0$, one can introduce the
\textit{generalized mean motion} and \textit{mean anomaly} as,
respectively,
\begin{align}
  \nu & =\frac{1}{\mu}\left(-2\mathscr{\mathscr{E}}\right)^{3/2},\label{eq:mm_gen}\\
  \mathcal{M} & =\nu(t-t_{p}),\nonumber 
\end{align}
where $t_{p}$ denotes the time of pericenter passage of $\Gamma$.

The angular separation between the vectors ${\bf g}$ and ${\bf r}$
defines the \textit{generalized true anomaly} $\theta$ which can be
obtained from the relations (analogous to the ones holding for the
classical true anomaly)
\[
\left\{
\begin{aligned}
  g\cos\theta & =\dfrac{c^{2}}{\mu r}-1,\\
  g\sin\theta & =\frac{cu}{\mu}.
\end{aligned}
\right.
\]
Similarly, a \textit{generalized eccentric anomaly $G$} can be defined
from the relations
\[
\left\{
\begin{aligned}
  g\cos G & =1-\frac{r}{\textrm{a}},\\
  g\sin G & =\frac{ru}{\sqrt{\textrm{\ensuremath{\mu}a}}}.
\end{aligned}
\right.
\]
The \textit{generalized Kepler's equation} takes the form
\[
\mathcal{M}=G-g\sin G.
\]

Let $\left\{ {\bf e}_{X},\,{\bf e}_{Y},\,{\bf e}_{h}\right\} $ be the
classical equinoctial basis introduced in \cite{Arsenault_1970}. The
angular separation between the vectors ${\bf e}_{X}$ and ${\bf g}$,
which both lie on the osculating orbital plane, is given by the
\textit{generalized longitude of periapsis}
\begin{equation}
  \Psi=L-\theta,\label{eq:gen_long_peri}
\end{equation}
where 
\begin{equation}
  L=\omega+\Omega+f\label{eq:tr_long}
\end{equation}
is the classical true longitude with $\omega$ and $\Omega$ denoting
the classical argument of pericenter and right ascension of the
ascending node, respectively. By means of the angle $\Psi$, one can
introduce the \textit{generalized} \textit{eccentric longitude} and
\textit{generalized mean longitudes} as, respectively,
\begin{align}
  \mathcal{\mathcal{K}} & =G+\Psi,\label{eq:ean_gen}\\
  \mathcal{\mathcal{L}} & =\mathcal{M}+\Psi.\label{eq:ml_gen}
\end{align}

\section{Generalized Equinoctial Orbital Elements (GEqOE)}

The generalized equinoctial orbital elements (GEqOE) as defined in
\cite{bau2021generalization} are constructed on the previously
described generalized quantities $g$ (Eq.~\ref{eq:ecc_gen}), $\nu$
(Eq.~\ref{eq:mm_gen}), $\Psi$ (Eq.~\ref{eq:gen_long_peri}), and
$\mathcal{L}$ (Eq.~\ref{eq:ml_gen}).  The six elements read:
\begin{align*}
  \nu, &  & p_{1} & =g\sin\Psi, & p_{2} & =g\cos\Psi,\\
  \mathcal{L}, &  & q_{1} & =\tan\frac{i}{2}\sin\Omega, & q_{2} & =\tan\frac{i}{2}\cos\Omega.
\end{align*}
The elements $p_{1}$, $p_{2}$ represent the projections of the
generalized eccentricity vector ${\bf g}$ along the equinoctial basis
unit vectors ${\bf e}_{Y}$ and ${\bf e}_{X}$, while the elements
$q_{1}$, $q_{2}$ are two of the classical equinoctial orbital elements
\cite{BC_1972}, where $i$ is the orbital inclination. Note that when
$\mathscr{U}=0$, the generalized equinoctial elements coincide with
the alternate equinoctial orbital elements (AEqOE) proposed in
\cite{Horwood2011} and further discussed in Section
\ref{sec:keplerian_motion}.

\subsection{Equations of motion}

Following \cite{bau2021generalization}, the time derivatives of the
GEqOE obey
\begin{align}
  \dot{\nu} & =-3\left(\frac{\nu}{\mu^{2}}\right)^{1/3}\dot{\mathscr{E}},
  \label{eq:nudot}\\
  \dot{p}_{1} & =p_{2}\biggl(\dfrac{h-c}{r^{2}}-\dfrac{\gamma}{h}F_{h}\biggr)+
  \frac{1}{c}\biggl(\frac{X}{\textrm{a}}+2p_{2}\biggr)(2\mathscr{U}-rF_{r})+
  \dfrac{1}{c^{2}}\left[Y(r+\varrho)+r^{2}p_{1}\right]\dot{\mathscr{E}},
  \label{eq:p1dot}\\
  \dot{p}_{2} & =p_{1}\biggl(\dfrac{\gamma}{h}F_{h}-\dfrac{h-c}{r^{2}}\biggr)-
  \frac{1}{c}\biggl(\frac{Y}{\textrm{a}}+2p_{1}\biggr)(2\mathscr{U}-rF_{r})+
  \dfrac{1}{c^{2}}\left[X\left(r+\varrho\right)+r^{2}p_{2}\right]\dot{\mathscr{E}},
  \label{eq:p2dot}\\
  \dot{\mathcal{L}} & =\nu+\dfrac{h-c}{r^{2}}-\dfrac{\gamma}{h}F_{h}+
  \frac{1}{c}\biggl[\frac{1}{\alpha}+\alpha\Bigl(1-\frac{r}{\textrm{a}}\Bigr)\biggr]
  (2\mathscr{U}-rF_{r})+\frac{ru\alpha}{\mu c}(r+\varrho)\dot{\mathscr{E}},
  \label{eq:Ldot}\\
  \dot{q}_{1} & =\frac{Y}{2h}F_{h}\left(1+q_{1}^{2}+q_{2}^{2}\right),
  \label{eq:q1dot}\\
  \dot{q}_{2} & =\frac{X}{2h}F_{h}\left(1+q_{1}^{2}+q_{2}^{2}\right),
  \label{eq:q2dot}
\end{align}
where 
\[
\dot{\mathscr{E}}=\frac{\partial\mathscr{U}}{\partial t}+uP_{r}+\frac{h}{r}P_{f}
\]
and 
\begin{equation}
  X=r\cos L,\qquad Y=r\sin L,
  \label{eq:XY}
\end{equation}
\begin{equation}
  \textrm{a}=\left(\frac{\mu}{\nu^{2}}\right)^{1/3},\qquad\varrho=\frac{c^{2}}{\mu},
  \label{eq:a_nu}
\end{equation}
\begin{equation}
  \gamma=Xq_{1}-Yq_{2},\qquad\alpha=\frac{1}{1+\beta},\qquad\beta=\sqrt{1-p_{1}^{2}-p_{2}^{2}}.
  \label{eq:g_al_be}
\end{equation}
Moreover, the terms $F_{r},F_{h},P_{r},P_{f}$ in the preceding
equations are the projections of $\mathbf{F}$ and $\mathbf{P}$ onto
the \textit{orbital} reference frame:
\begin{equation}
  F_{r}={\bf F}\cdot{\bf e}_{r},\quad F_{h}={\bf F}\cdot{\bf e}_{h},\quad P_{r}=
  {\bf P}\cdot{\bf e}_{r},\quad P_{f}={\bf P}\cdot{\bf e}_{f},
  \label{eq:FPprojs}
\end{equation}
where the corresponding unit vectors can be conveniently obtained as
\[
  {\bf e}_{r}=\frac{1}{r}(X{\bf e}_{X}+Y{\bf e}_{Y}),\qquad{\bf e}_{f}=
  \frac{1}{r}(X{\bf e}_{Y}-Y{\bf e}_{X}),\qquad{\bf e}_{h}={\bf e}_{r}\times{\bf e}_{f}.
\]

Given the initial position ${\bf r}_{0}$ and velocity $\dot{{\bf
    r}}_{0}$ at some time $t_{0}$ expressed in a suitable inertial
reference frame $\Sigma$, the motion can be propagated to the epoch
$t\ne t_{0}$ by obtaining the corresponding GEqOE initial conditions
(see the conversion formulas in \cite{bau2021generalization},
Sect.~3), integrating Eqs.~(\ref{eq:nudot}--\ref{eq:q2dot}) and
converting back the state expressed in GEqOE at time $t$ to Cartesian
variables (see the conversion formulas in
\cite{bau2021generalization}, Sect.~4).

\section{Linear Covariance Propagation in GEqOE\label{sec:LCP}}

In the following, the linear propagation, by use of the propagated
state transition matrix, of the covariance matrix of the GEqOE set is
dealt with in details.

Note that in general any of the methods described for pdf propagation
in the Introduction can be combined with GEqOE, boosting their
performance or reducing their computational cost. Since the objective
here is to investigate the suitability of GEqOE for orbital
uncertainty propagation, nonlinear methods are out of the scope of
this work.

\subsection{State Transition Matrix Propagation}

Equations \eqref{eq:nudot}--\eqref{eq:q2dot} can be written as
\begin{equation}
  \dot{\bf y}={\bf f}({\bf y},t),
  \label{eq:dydt}
\end{equation}
where ${\bf y}=(\nu,\,p_{1},\,p_{2},\,\mathcal{L},\,q_{1},\,q_{2})^{T}$
denotes the state vector in GEqOE.

An orbit ${\bf y}(t)$ close to a reference orbit ${\bf y}_{*}(t)$ can
be propagated linearly in time from an initial epoch $t_{0}$ as
\[
\mathbf{y}(t)\approx\mathbf{y}_{*}(t)+\Phi(t,t_0)(\mathbf{y}(t_{0})-\mathbf{y}_{*}(t_{0})),
\]
where the reference orbit state transition matrix $\Phi$ is the
solution of the linear Cauchy problem
\begin{equation}
  \left\{
  \begin{aligned}
    & \frac{\partial\Phi}{\partial t}=\frac{\partial{\bf f}}
    {\partial{\bf y}}({\bf y}_{*}(t),t)\,\Phi,\\
    & \Phi(t_{0},t_{0})=I,
  \end{aligned}
  \right.
  \label{eq:VE}
\end{equation}
with $I$ the $6\times6$ identity matrix. The matrix $\partial{\bf
  f}/\partial{\bf y}$ can be computed analytically for the GEqOE as
described in Appendix \ref{sec:dfdy}.

\subsection{Linear Covariance Propagation and Nonlinear Mapping into Cartesian
Coordinates}

The state transition matrix $\Phi(t,t_{0})$ can be employed for the
linear covariance matrix propagation in GEqOE space, $P_{{\bf y}}$, as
\[
P_{{\bf y}}(t)=\Phi(t,t_{0})\,P_{{\bf y}}(t_{0})\,\Phi^{T}(t,t_{0}).
\]
In practice, the GEqOE covariance matrix at $t_{0}$ can be obtained
starting from the corresponding covariance matrix, $P_{\bf x}$, in
another set of coordinates $\mathbf{x}$ (e.g. Cartesian elements) by
the linear mapping
\begin{equation}
  P_{{\bf y}}(t_{0})={\rm J}(t_{0})\,P_{{\bf x}}(t_{0})\,{\rm J}^{T}(t_{0}),\qquad{\rm J}=
  \frac{\partial{\bf y}}{\partial{\bf x}}.
  \label{eq:Gx2Gy}
\end{equation}
The Jacobian matrix ${\rm J}$ for the case of Cartesian coordinates is
given in \cite{bau2021generalization}, Sect. 6.2.

\begin{figure}[H]
  \begin{centering}
    \includegraphics[width=10cm]{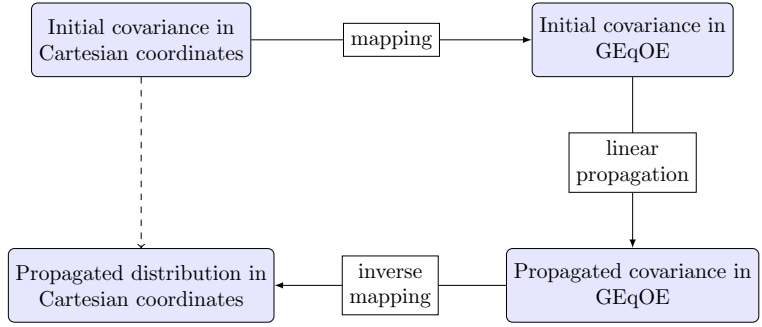}
    \par
  \end{centering}
  \caption{\label{UP_prop}Uncertainty propagation scheme.}
\end{figure}

After linearly propagating $P_{{\bf y}}(t)$ until epoch $t$, the
uncertainty cloud can be mapped back to Cartesian coordinates by a
full \textit{nonlinear} transformation employing the element
conversion equations in \cite{bau2021generalization}, Sect. 4. The
whole scheme, which provides a very efficient method for propagating
an initial uncertainty cloud is depicted in Figure \ref{UP_prop}. It
will be shown, in the reminder of this article, that this procedure is
particularly resilient against perturbation-driven nonlinear effects
hence preserving the UR of the distribution for a longer timespan
compared to other methods.

\section{Mitigation of Nonlinear Effects}

In this section a simple mathematical explanation is provided to
demonstrate the ability of the newly proposed uncertainty propagation
method to minimize UR losses.

\subsection{Keplerian motion}
\label{sec:keplerian_motion}

As detailed in \cite{Horwood2011}, when an element-based
representation, such as classical equinoctial elements, is employed,
five of the six coordinates evolve linearly (they are actually
constant) under Keplerian orbital motion while the evolution of the
time-element coordinate (the mean longitude, in the case of classical
equinoctial elements) is not exactly linear with respect to the other
state elements.

After indicating with $a_{*}$ the initial (reference) semi-major axis,
$\ell$ the mean longitude, and assuming an initial uncertainty in the
semi-major axis $\varepsilon_{0}^{a}$, the time evolution of the
uncertainty-affected mean longitude for a Keplerian orbit obeys
\begin{equation}
  \ell(t)=\ell_{0}+\sqrt{\frac{\mu}{(a_{*}+\varepsilon_{0}^{a})^{3}}}(t-t_{0}),
  \label{eq:error_affected mean longitude}
\end{equation}
which yields a \textit{secular} growth of the mean longitude
uncertainty:
\begin{equation}
  \varepsilon^{\ell}(t)=\ell(t)-\left.\ell(t)\right|_{\varepsilon_{0}^{a}=0}=
  \biggl(-\frac{3}{2a_{*}}\varepsilon_{0}^{a}+\frac{15}{8a_{*}^2}(\varepsilon_{0}^{a})^{2}
  +O\bigl((\varepsilon_{0}^{a})^{3}\bigr)\biggr)n_{*}(t-t_{0}),
  \label{eq:mean longitude error}
\end{equation}
where $n_{*}(t-t_{0})$ is the accumulated mean longitude difference
for the nominal (uncertainty free) orbit since the initial epoch and
$n_{*}$ is the mean motion of that nominal orbit.

Equation~(\ref{eq:mean longitude error}) shows that a linear
propagation of a pdf expressed in classical equinoctial elements will
be imprecise as a result of the truncation error, growing linearly in
time, in the propagation of the mean longitude.

As originally proposed by \cite{Horwood2011}, the classical
equinoctial elements can be improved by employing the mean motion as
an element in place of the semi-major axis. The resulting elements,
coined ``alternative equinoctial orbital elements'' (AEqOE), are
effective against the above-mentioned nonlinear truncation error.

If the pdf is expressed with respect to AEqOE, an initial uncertainty
in the mean motion variable $\varepsilon_{0}^{n}$ corresponds to
\begin{equation}
  \ell(t)=\ell_{0}+(n_{*}+\varepsilon_{0}^{n})(t-t_{0}),
  \label{eq:error_affected_ML_AEqOE}
\end{equation}
\begin{equation}
  \varepsilon^{\ell}(t)=\varepsilon_{0}^{n}(t-t_{0}),
  \label{eq:ML_error_AEqOE}
\end{equation}
providing a time evolution of the mean longitude uncertainty that is
perfectly linear with respect to the initial mean motion uncertainty.

\subsection{J$_2$-induced Secular Perturbations}

Let us now consider the case of a perturbed orbital motion whose
perturbing force is fully derivable from a potential. In particular,
let us consider the effect of the J$_2$ term of the gravity field
harmonics. Analogously to the previous section, we assume an initial
semi-major axis uncertainty $\varepsilon_{0}^{a}$.

If the orbital motion is represented in AEqOE, the assumed semi-major
axis uncertainty corresponds to a mean motion uncertainty
\[
\varepsilon_{0}^{n}=-\frac{3}{2}\sqrt{\frac{\mu}{a^{5}}}\,\varepsilon_{0}^{a}+
O\bigl((\varepsilon_{0}^{a})^{2}\bigr).
\]
On the other hand, if the orbital motion is represented in GEqOE, it
is possible to show that the assumed semi-major axis uncertainty
corresponds to a generalized mean motion uncertainty
\[
\varepsilon_{0}^{\nu}=-\frac{3}{2}\sqrt{\frac{\mu}{a^{5}}}\,\varepsilon_{0}^{a}+O_{2}
\bigl(\varepsilon_{0}^{a},J_{2}\bigr).
\]
The previous uncertainties $\varepsilon_{0}^{n}$,
$\varepsilon_{0}^{\nu}$ are equal to first order in
$\varepsilon_{0}^{a}$.  However, their nonlinear contribution to the
evolution of the, respectively, mean motion and generalized mean
motion errors, will be here shown to be different.

Following \cite{battin1999introduction}, the expression of the
disturbing function after averaging over one orbital period reads
\[
\overline{\mathscr{R}}=\frac{J_{2}R^{2}n^{2}}{4(1-e^{2})^{3/2}}(2-3\sin^{2}i),
\]
where, without complicating the notation, $e$, $i$, and $n$ stand here
for the mean values of eccentricity, inclination, and mean motion,
respectively. The mean rate of the mean longitude can be derived from
Lagrange's planetary equations for $-n\tau$ (with $\tau$ the time of
pericenter passage) wherein the disturbing function is replaced by
$\overline{\mathscr{R}}$ (see \cite{battin1999introduction}):
\begin{equation}
  \dot{\ell}=n+\frac{J_{2}R^{2}n^{7/3}}{\mu^{2/3}}\lambda(e,i),
  \label{eq:delle}
\end{equation}
with 
\[
\lambda=\frac{3}{2(1-e^{2})^{2}}\Bigl[\bigl(1+\sqrt{1-e^{2}}\bigr)
\Bigl(1-\frac{3}{2}\sin^{2}i\Bigr)+\cos^{2}i-\cos i\Bigr].
\]
Note that $n$ and $i$ are constant (there are no secular nor
long-period variations of these elements under J$_2$
\cite{brouwer1959solution}) while $e$ is constant after neglecting
J$_2$-induced long-period effects.

Let us consider, for simplicity, the case of a circular equatorial
orbit ($i=0$, $e=0$) and introduce the notation
\begin{equation}
  A=\frac{3R^{2}}{2\mu^{2/3}}.
  \label{eq:A}
\end{equation}
Integration of equation (\ref{eq:delle}) yields 
\begin{equation}
  \ell=\ell_{0}+n(t-t_{0})\bigl(1+2n^{4/3}J_{2}A\bigr).
  \label{eq:elle_t}
\end{equation}
After writing the mean motion as the sum of its nominal value and the
associated error:
\[
n=n_{*}+\varepsilon_{0}^{n},
\]
and by developing $n^{4/3}$ (in Eq.~\ref{eq:elle_t}) in Taylor series
with respect to $\varepsilon_{0}^{n}$, the mean longitude error can be
obtained as
\begin{equation}
  \varepsilon^{\ell}=(t-t_{0})\biggl[\biggl(1+\frac{14}{3}J_{2}An_{*}^{4/3}\biggr)\varepsilon_{0}^{n}+
    \biggl(\frac{28}{9}J_{2}An_{*}^{1/3}\biggr)(\varepsilon_{0}^{n})^{2}+
    O\bigl((\varepsilon_{0}^{n})^{3}\bigr)\biggr],
  \label{eq:mean_lon_uncertainty}
\end{equation}
which is characterized by a secular growth with a \textit{nonlinear}
dependency on the mean motion error due to the J$_2$ term.

A similar analysis can be developed for a formulation based on GEqOE.
When the GEqOE are employed, the generalized mean motion state
variable $\nu$ is a constant in the J$_2$-only perturbed two-body
problem.  Moreover, for an equatorial orbit the potential energy
associated with the J$_2$ term yields
\begin{equation}
  \mathscr{U}=-\frac{J_{2}\mu R^{2}}{2r^{3}},
  \label{eq:UJ2}
\end{equation}
and the time derivative of the generalized mean longitude becomes
(Eq.~\ref{eq:Ldot})
\begin{equation}
  \dot{\mathcal{L}}=\nu_{0}+\dfrac{h-c}{r^{2}}+\frac{J_{2}\mu R^{2}}{2cr^{3}}
  \biggl[\frac{1}{\alpha}+\alpha\Bigl(1-\frac{r}{\textrm{a}_{0}}\Bigr)\biggr],
  \label{eq:dL_J2}
\end{equation}
where $\alpha$ has been defined in \eqref{eq:g_al_be} and $\nu_{0}$,
$\textrm{a}_{0}$ are the constant values taken by the generalized mean
motion and semi-major axis along the solutions.

Noting from Eqs.~\eqref{eq:am_gen},~\eqref{eq:UJ2} that
\begin{equation}
  \frac{h-c}{r^{2}}=\frac{J_{2}\mu R^{2}}{2cr^{3}}+O(J_{2}^{2}),
  \label{eq:hmc}
\end{equation}
the secular variation (to first order in J$_{2}$) in generalized mean
longitude for a circular equatorial orbit is given by (see Appendix
\ref{sec:secL_J2})
\begin{equation}
  \mathcal{L}=\mathcal{L}_{0}+\nu_{0}(t-t_{0})\bigl(1+J_{2}A\nu_{0}^{4/3}\bigr).
  \label{eq:elleg_t}
\end{equation}
Writing the generalized mean motion as the sum of its (constant)
nominal value and the associated error:
\[
\nu_{0}=\nu_{*}+\varepsilon_{0}^{\nu},
\]
and expanding $\nu_0^{4/3}$ (in Eq.~\ref{eq:elleg_t}) in Taylor series
with respect to $\varepsilon_{0}^{\nu}$ yields the generalized mean
longitude error
\begin{equation}
  \varepsilon^{\mathcal{L}}=(t-t_{0})\biggl[\biggl(1+\frac{7}{3}J_{2}A\nu_{*}^{4/3}\biggr)\varepsilon_{0}^{\nu}+
  \biggl(\frac{14}{9}J_{2}A\nu_{*}^{1/3}\biggr)(\varepsilon_{0}^{\nu})^{2}+
  O\bigl((\varepsilon_{0}^{\nu})^{3}\bigr)\biggr].
  \label{eq:gen_mean_long_uncertainty}
\end{equation}

After comparing Eq.~(\ref{eq:mean_lon_uncertainty}) with
Eq.~(\ref{eq:gen_mean_long_uncertainty}) there appears to be a
reduction by a factor of two for the nonlinear dependency on the mean
motion error. This means that the use of GEqOE has absorbed half of
the J$_{2}$ nonlinear secular effect on the mean longitude. Indeed,
simple tests for circular equatorial orbits have confirmed an
improvement of UR by a factor very close to two.  Extending the
present analytical treatment to the much more complex case of
non-circular and non-equatorial orbits is out of the scope of the
present article.

\section{Results}

To assess the efficiency of the orbital uncertainty propagation using
the GEqOE set, a Cram\'er-von Mises (CvM) test of the Mahalanobis
distance distribution is used following the recent work of Aristoff et
al.  \cite{AHA_2021}. This test evaluates whether a covariance matrix,
based on a chosen set of coordinates, is likely to represent the true
covariance. For convenience, the process is briefly described in
Algorithm \ref{CvM_algo}. More details can be found in
\cite{AHA_2021}.

The first step is to propagate the covariance matrix in the
coordinates set of interest. A linear propagation of that matrix can
be adopted for simplicity.

Subsequently, a sufficiently large set ($N=10000$ for these results)
of orbital states sampled from the initial covariance are propagated
with a high-fidelity orbital dynamics model in Cartesian coordinates
and converted into the set of variables being evaluated. For each
epoch, the Mahalanobis distance $\mathscr{\mathfrak{m}}$ of each
sample is calculated using the linearly propagated covariance and the
propagated state of the sample. If the true pdf is Gaussian, then the
Mahalanobis distance will follow a chi-squared distribution with 6
degrees of freedom, whose cumulative distribution function (cdf) is
\begin{equation}
  F(z)=1-\frac{1}{8}\exp\Bigl({-\frac{z}{2}}(z^{2}+4z+8)\Bigr).
\end{equation}
The CvM test statistics is calculated by comparing this cdf with the
empirical cumulative distribution function of the true distribution.
If the CvM test is satisfied, the two cdf are found to agree and one
can say that the covariance is realistic.

\begin{algorithm}
  \setstretch{1.35} \caption{Cram\'er-von Mises test}

  \label{CvM_algo}

  \begin{algorithmic}[1]
    \State{$P_{\mathbf{y}}(t)=\Phi(t,t_{0})\,P_{\mathbf{y}}(t_{0})\,\Phi^{T}(t,t_{0})$}
    \Comment{covariance matrix linear propagation}
    \For{$i=1,\ldots,N$}
    \State{$\mathbf{x}_{i}(t_{0})$} 
    \Comment{initial covariance sampling}
    \State{$\mathbf{x}_{i}(t)$}
    \Comment{true orbit of the sample propagation}
    \State{$\mathbf{y}_{i}(t)=\mathbf{y}(\mathbf{x}_{i}(t),t)$}
    \Comment{conversion to state variables being tested}
    \EndFor
    \State{$\boldsymbol{\mu}_{\mathbf{y}}(t)=\frac{1}{N}\sum_{i}^{N}\mathbf{y}_{i}(t)$}
    \Comment{true mean orbit}
    \For{$i=1,\ldots,N$}
    \State{$\mathfrak{M}_{i}(\mathbf{y}_{i};
      \boldsymbol{\mu}_{\mathbf{y}},P_{\mathbf{y}})=(\mathbf{y}_{i}-\boldsymbol{\mu}_{\mathbf{y}})^{T}
      P_{\mathbf{y}}^{-1}(\mathbf{y}_{i}-\boldsymbol{\mu}_{\mathbf{y}})$}
    \Comment{Mahalanobis distance}
    \EndFor
    \State{$Q=\frac{1}{12N}$}
    \Comment{CvM test statistics initialization}
    \For{$j=1,\ldots,N$}
    \Comment{in increasing order of the Mahalanobis distance}
    \State{$Q=Q+\left(\frac{2j-1}{2N}-F(\mathfrak{M}_{j})\right)^{2}$}
    \Comment{$F$ is the cdf of a 6D chi-squared}
    \EndFor
    \If{$Q<Q^{*}\simeq1.16$}
    \State{covariance is realistic}
    \EndIf
  \end{algorithmic} 
\end{algorithm}

The Cram\'er-Von Mises test is applied to the proposed GEqOE
formulation.  For comparison, results for the J2EqOE elements recently
proposed by Aristoff et al. \cite{AHA_2021} and for the alternative
equinoctial elements (AEqOE) \cite{Horwood2011} are also computed.

The following three different test cases are analyzed:
\begin{itemize}
\item[] Case 1)$\quad$ LEO from \cite{AHA_2021}, 
\item[] Case 2)$\quad$ High-Earth orbit (HEO) from \cite{AHA_2021}, 
\item[] Case 3)$\quad$ super Geostationary transfer orbit (super-GTO). 
\end{itemize}
For each of these three test cases two scenarios are investigated: (a)
a ballistic scenario and (b) a constant low-thrust tangential
acceleration scenario. The low-thrust-perturbed scenarios are defined
based on mega-constellation LEO satellites (Case 1) and all-electric
satellites maneuvered to GEO (Cases 2 and 3). The employed control law
sets the thrust vector tangent to the nominal trajectory and with
constant magnitude. Note that in practical applications new
measurements are performed leading to a continuous update of the
covariance statistics readily available to the owner/operator of a
specific satellite and possibly distributed to other parties as
well. Nevertheless, it is still relevant to study how such covariance
would evolve without considering new measurements in case these
updates were not available.

For the LEO case, the thrust magnitude and the spacecraft mass are set
to 15\,mN and 260\,kg, respectively, based on realistic estimates for
Starlink satellites\footnote{A nominal 260\,kg wet mass has been
  assumed based on NASA Space Science Data Coordinate Archive
  (\url{https://nssdc.gsfc.nasa.gov/nmc/spacecraft/display.action?id=2019-074D}).
  In addition, a roughly estimated 15\,mN thrust after fitting TLE
  data of Starlink satellites during their spiral up phase has been
  considered.}. For the HEO and super-GTO, 165\,mN and 2200\,kg were
used in line with published figures for the Eutelsat 115 West B
satellite\footnote{Based on TLE data, EutelSat 115 West B was boosted
  from a $\approx\,$70000\,km apogee super-GTO to a GEO from March to
  October 2015. The satellite employed a XIPS-25 propulsion system of
  $\approx\,$165\,mN maximum thrust capability with an estimated wet
  mass of 2200\,kg \cite{abbasrezaee2019conceptual}.}.

The initial orbital conditions are shown in
Table~\ref{table_initial_conditions}, where all the elements are
expressed in the Earth-centered J2000 frame.  The initial epoch is
2021 October 20 00:00:00 TDB and the initial covariance in equinoctial
orbital elements (EqOE)\footnote{Note that $P_{1}$, $P_{2}$, $\ell$
  are the osculating orbital elements corresponding to $p_{1}$,
  $p_{2}$, $\mathcal{L}$ of GEqOE, and coincide with them if
  $\mathscr{U}=0$.} is given in Table~\ref{table_initial_covariance}.

\begin{table}
\caption{Initial orbital elements for Cases 1, 2, 3. Angles are in degrees.}
\vspace{0.3cm}
 
\global\long\def\arraystretch{1.5}%
 
\begin{centering}
  \label{table_initial_conditions} 
  \par\end{centering}
  \centering{}%
  \begin{tabular}{lrld{-1}cd{-1}d{-1}}
    \hline 
    & \mc{$a$ (km)}  & \mc{$e$}  & \mc{$i$}  & \mc{$\Omega$}  & \mc{$\omega$}  & \mc{$M$}\tabularnewline
    \hline 
    1)  & 7136.6  & 0.00949  & 72.9  & 116  & 57.7  & 105.5\tabularnewline
    2)  & 26628.1  & 0.742  & 63.4  & 120  & 0  & 144\tabularnewline
    3)  & 38200.0  & 0.8167539267  & 25  & 120  & 0  & 0\tabularnewline
    \hline 
  \end{tabular}
\end{table}

\begin{table}
  \caption{Initial covariance in EqOE for Cases 1, 2, 3.}
  \vspace{0.3cm}
  
  \global\long\def\arraystretch{1.5}%
  
  \begin{centering}
    \label{table_initial_covariance} 
    \par\end{centering}
    \centering{}%
    \begin{tabular}{ld{-1}ccccc}
      \hline 
      & \mc{$\sigma_{a}$ (km)}  & $\sigma_{P_{1}}$  & $\sigma_{P_{2}}$  & $\sigma_{q_{1}}$
      & $\sigma_{q_{2}}$  & $\sigma_{\ell}$ (deg)\tabularnewline
      \hline 
      1)  & 20  & $10^{-3}$  & $10^{-3}$  & $10^{-3}$  & $10^{-3}$  & $10^{-2}$\tabularnewline
      2)  & 2  & $10^{-4}$  & $10^{-4}$  & $10^{-4}$  & $10^{-4}$  & $\frac{7}{900}$\tabularnewline
      3)  & 2  & $10^{-4}$  & $10^{-4}$  & $10^{-4}$  & $10^{-4}$  & $\frac{7}{900}$\tabularnewline
      \hline
    \end{tabular}
\end{table}

The sampled states are propagated using Matlab's ode113
(Adams--Bashforth--Moulton predictor corrector method with variable
order between 1 and 13), while the predictions of the nominal state
and the associated covariance matrix are carried out with a
Runge-Kutta method of 4th order and a time step close to 60
seconds. To this end, the state transition matrix in GEqOE is computed
using the variational equations provided in Appendix~\ref{sec:dfdy}.
Instead, for AEqOE and J2EqOE the state transition matrix is computed
from that in GEqOE using the Jacobian of the corresponding
transformation.

All scenarios consider luni-solar gravitational perturbations as
forces not included in the potential energy defining the GEqOE, as
recommended in \cite{bau2021generalization}. The position of these
celestial bodies are obtained from the JPL DE430 ephemeris. The Earth
gravitational field is modeled following Grace gravity model 05 GGM05C
\cite{ries2016development} truncated to the 8th degree and order, and
all its terms except for the point mass potential are included in the
potential energy $\mathscr{U}$.  The Earth-centered Earth-fixed
coordinate system (ECEF) is set as ITRF93.  The calculation of the
Earth gravitational potential, its gradient and Hessian in ECEF are
performed following the method of Cunningham and Metris
\cite{cunningham1970computation,metris1998derivatives}.  The details
about the procedure used to include the Earth gravitational potential
in the GEqOE formulation are described in Appendix~\ref{sec:shp_der}.

Figure~\ref{fig:LEO} shows the CvM test statistics as a function of
time for the no-thrust scenario of Case 1. This case was analyzed by
Aristoff et al. \cite{AHA_2021} and features a relatively-low initial
orbit accuracy (see Table~\ref{table_initial_covariance}).  As
Aristoff et al. showed, the J2EqOE formulation can represent the real
orbit uncertainty longer than simpler methods that do not account for
the J$_{2}$ perturbation like AEqOE. In particular, the CvM test fails
before 6 orbital periods in J2EqOE, and before 2 orbital periods for
AEqOE. The new proposed GEqOE formulation can further extend the
covariance realism to almost 8 orbits.

\begin{figure}[H]
  \begin{centering}
    \includegraphics[width=0.8\textwidth]{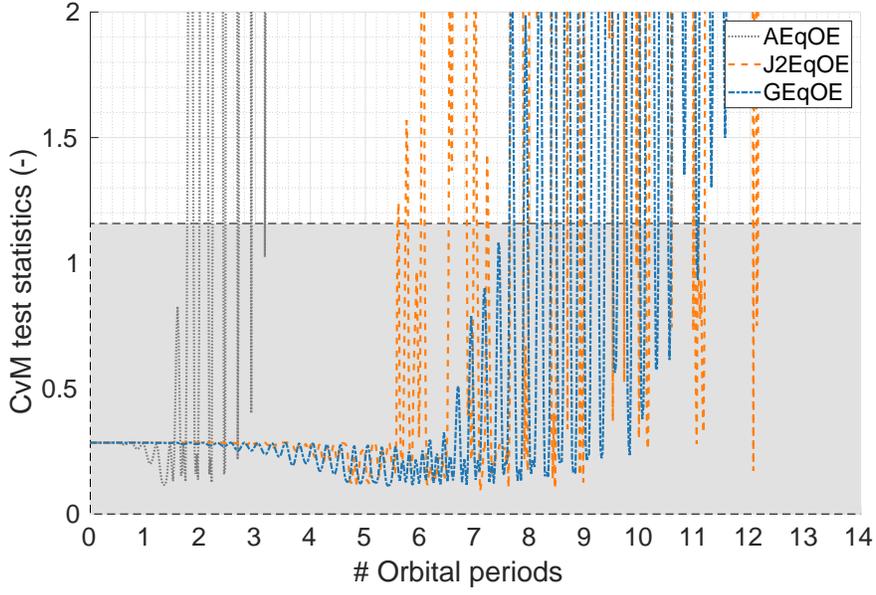} 
    \par
  \end{centering}
  \caption{\label{fig:LEO}CvM test statistics for Case 1 (ballistic).}
\end{figure}

\begin{figure}[H]
  \begin{centering}
    \includegraphics[width=0.8\textwidth]{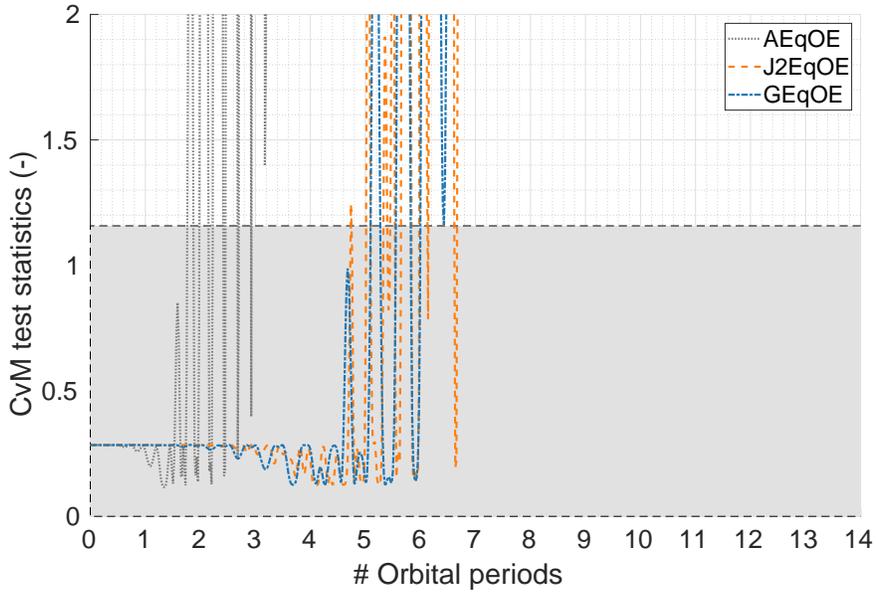} 
    \par
  \end{centering}
  \caption{\label{fig:LEO-lowthrust}CvM test statistics for Case 1 (low-thrust).}
\end{figure}

The UR improvement of the GEqOE formulation can be partly reduced if a
significant acceleration that cannot be included in the potential
energy $\mathscr{U}$ perturbs the trajectory. Nevertheless, in spite
of the fact that such reduction is more significant for GEqOE compared
to the AEqOE and J2EqOE formulations (see
Fig.~\ref{fig:LEO-lowthrust}), the former is still able to stay
Gaussian longer than the other two.

\begin{figure}[H]
  \begin{centering}
    \includegraphics[width=0.8\textwidth]{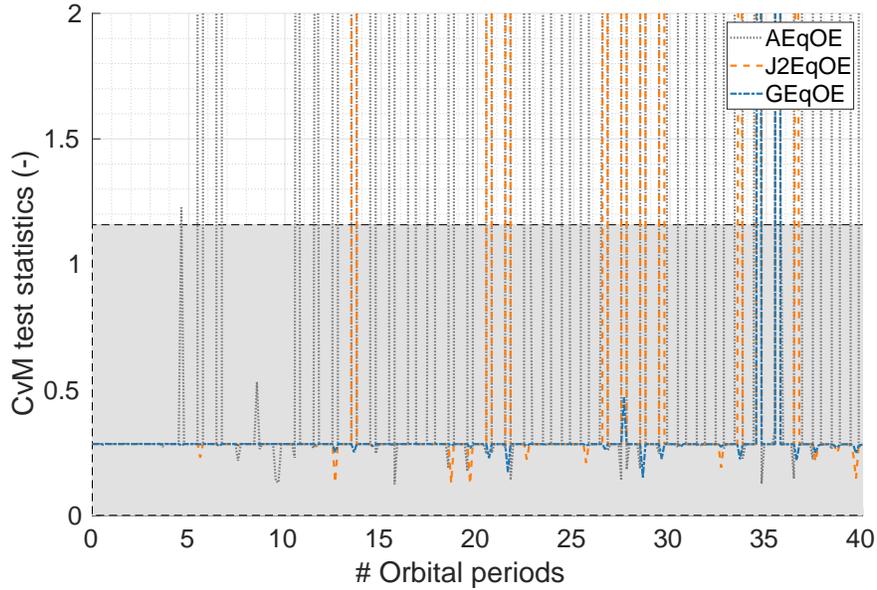} 
    \par
  \end{centering}
  \caption{\label{fig:HEO}CvM test statistics for Case 2 (ballistic).}
\end{figure}

Ballistic Case 2 was also analyzed by Aristoff et al. \cite{AHA_2021},
and it was pointed out that the realism was lost near the periapses.
This was confirmed by our simulations as seen in Fig. \ref{fig:HEO}.
AEqOE loses realism after a few orbits, while it takes more than 10
revolutions for J2EqOE. Remarkably, using a GEqOE formulation, the UR
can be prolonged for more than 30 orbits in this case.

\begin{figure}[H]
  \begin{centering}
    \includegraphics[width=0.8\textwidth]{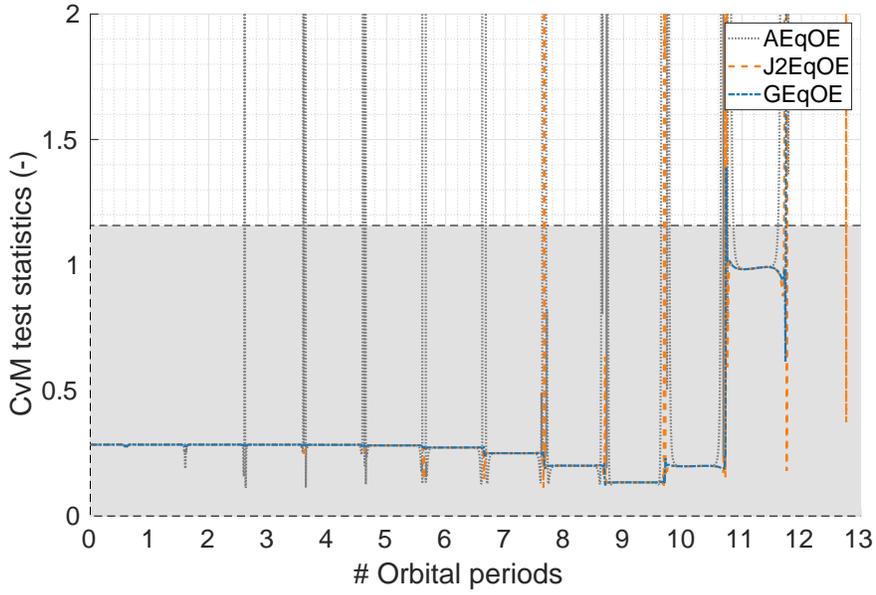} 
    \par
  \end{centering}
  \caption{\label{fig:HEO_lowthrust}CvM test statistics for Case 2 (low-thrust).}
\end{figure}

Figure~\ref{fig:HEO_lowthrust} shows the effect of adding a tangential
thrust to Case 2. The three formulations can only accurately describe
the real uncertainty for shorter periods of time, but, still, the
GEqOE formulation increases the realism by about 4 orbits compared to
the J2EqOE one.

\begin{figure}[H]
  \begin{centering}
    \includegraphics[width=0.8\textwidth]{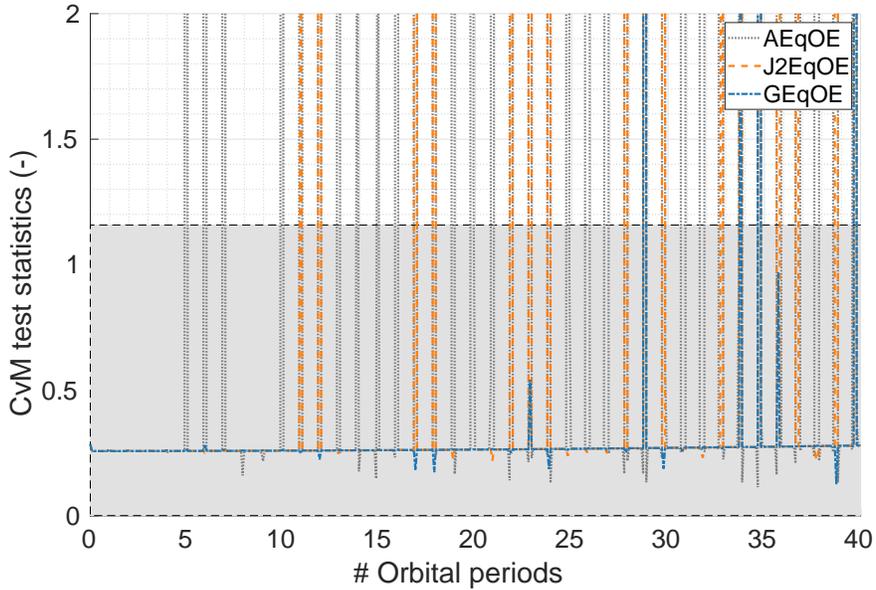} 
    \par
  \end{centering}
  \caption{\label{fig:GTO}CvM test statistics for Case 3 (ballistic).}
\end{figure}

\begin{figure}[H]
  \begin{centering}
    \includegraphics[width=0.8\textwidth]{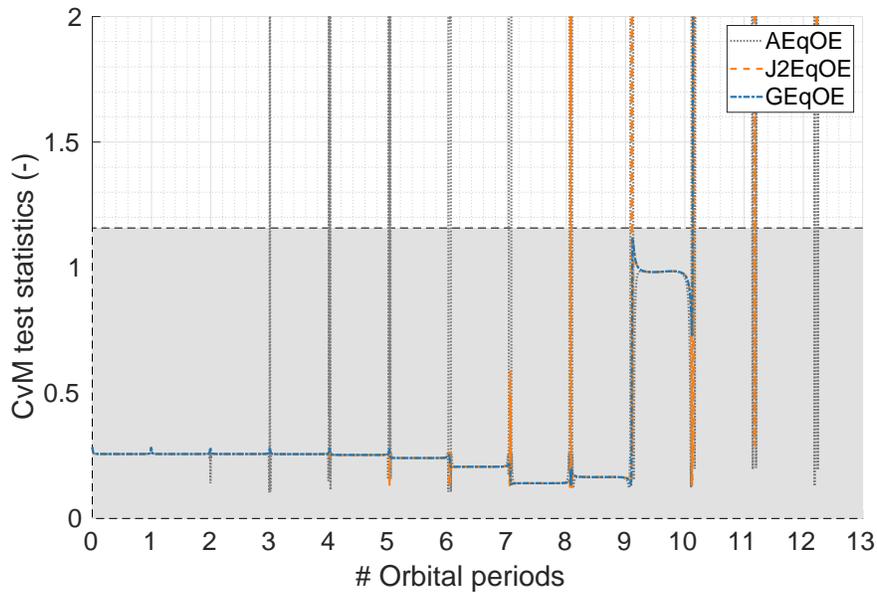} 
    \par
  \end{centering}
  \caption{\label{fig:GTO_lowthrust}CvM test statistics for Case 3 (low-thrust).}
\end{figure}

The results for Case 3 are shown in Figs.~\ref{fig:GTO} and
\ref{fig:GTO_lowthrust} for the ballistic and low-thrust scenarios,
respectively. The results are similar to the previous case and GEqOE
shows better performance than the other methods. In particular, the
covariance realism is conserved for about 29 orbits compared to the 11
orbits of J2EqOE in the absence of thrust. If the propulsion system is
active through the propagation period, GEqOE (10 orbits) can still
better predict the distribution than J2EqOE (8 orbits).

The six scenarios are summarized in Table~\ref{tab:summary}. The last
column displays the UR obtained by including in the potential energy
only the J$_2$ term of GEqOE. This result, not shown in the plots for
clarity, highlights the benefit of embedding higher order terms of the
geopotential into the definition of GEqOE especially for high
eccentricity orbits where the improvement in realism can be dramatic.

\begin{table}
  \caption{\label{tab:summary}Number of revolutions for the CvM test before
    failure.}
  \vspace{0.3cm}
  
  \global\long\def\arraystretch{1.5}%
  
  \centering{}%
  \begin{tabular}{ld{-1}d{-1}d{-1}d{-1}}
    \hline 
    Case/Scenario  & \mc{AEqOE}  & \mc{J2EqOE}  & \mc{GEqOE}  & \mc{GEqOE(J$_2$)} \tabularnewline
    \hline 
    1/a (LEO, ballistic)  & 1.77  & 5.60  & 7.62  & 5.67 \tabularnewline
    1/b (LEO, low-thrust)  & 1.77  & 4.74  & 5.11  & 4.66 \tabularnewline
    2/a (HEO, ballistic)  & 4.62  & 13.60  & 34.61  & 13.60\tabularnewline
    2/b (HEO, low-thrust)  & 2.60  & 7.66  & 10.73  & 7.66\tabularnewline
    3/a (super-GTO, ballistic)  & 4.98  & 10.93  & 28.79  & 10.93\tabularnewline
    3/b (super-GTO, low-thrust)  & 3.00  & 8.07  & 10.13  & 7.05 \tabularnewline
    \hline 
  \end{tabular}
\end{table}

\section{Conclusions}

A new linear uncertainty propagation scheme based on a set of
generalized equinoctial orbital elements (GEqOE) has been proposed and
shown to be superior, in terms of uncertainty realism (UR), to all
other linear propagation methods proposed so far. A UR improvement,
computed with a Mahalanobis distance Cram\'er-von Mises test, of more
than at least 36\% is obtained, compared to the use of J2 equinoctial
orbital elements (J2EqOE), for different classes of ballistic Earth
orbits. A key result of the article concerns the influence of higher
order geopotential harmonics in degrading UR and the possibility of
drastically reducing this effect by the use of GEqOE. It is seen that
when harmonics terms of higher order than J$_{2}$ are embedded in the
definition of the GEqOE, a considerable improvement in UR is observed
for all Earth-orbiting scenarios analyzed. The improvement becomes
dramatic when highly eccentric orbits are considered. Finally, the
analysis of the impact of a continuous tangential low-thrust
acceleration on the UR of different classes of Earth orbits
considering full-thrust capability of state-of-the-art electric
propulsion systems suggests a small, although not negligible reduction
in UR for all cases. Nevertheless, GEqOE retain their advantage over
competing sets of elements even for low-thrust-perturbed orbits.

While the GEqOE uncertainty propagation scheme investigated in this
article employs a linear model for maximum computation efficiency,
nonlinear techniques can be applied to the same set of elements to
further improve the UR at the expense of a reduced computation
efficiency.

\section*{Acknowledgments}

This work was supported by MINECO/AEI and FEDER/EU under Project
PID2020-112576GB-C21. The authors thank the MINECO/AEI of Spain for
their financial support.

Alicia Mart\'inez-Cacho, of Universidad Polit\'ecnica de Madrid (UPM),
was supported by a PhD grant under UPM ``Programa Propio''.

The views expressed are those of the authors and do not necessarily
represent the views of the ispace-inc.

\bibliographystyle{AAS_publication}
\bibliography{bibliography}

\begin{thebibliography}{10}

\bibitem{Maybeck1982vol2}
P.~S. Maybeck, {\em Stochastic models, estimation, and control Volume 2},
  Vol.~141-2 of {\em Mathematics in Science and Engineering}.
\newblock New York: Academic press, 1982.

\bibitem{Armellin2010}
R.~Armellin, P.~Di~Lizia, F.~Bernelli-Zazzera, and M.~Berz, ``Asteroid close
  encounters characterization using differential algebra: the case of
  {A}pophis,''  {\em Celestial Mechanics and Dynamical Astronomy}, Vol.~107,
  No.~4, 2010, pp.~451--470, 10.1007/s10569-010-9283-5.

\bibitem{Park2006}
R.~S. Park and D.~J. Scheeres, ``Nonlinear mapping of {G}aussian statistics:
  theory and applications to spacecraft trajectory design,''  {\em Journal of
  Guidance, Control, and Dynamics}, Vol.~29, No.~6, 2006, pp.~1367--1375,
  10.2514/1.2017.

\bibitem{roa2021reduced}
J.~Roa and R.~S. Park, ``Reduced Nonlinear Model for Orbit Uncertainty
  Propagation and Estimation,''  {\em Journal of Guidance, Control, and
  Dynamics}, 2021, pp.~1--15, 10.2514/1.G005519.

\bibitem{Giza2009}
D.~Giza, P.~Singla, and M.~Jah, ``An Approach for Nonlinear Uncertainty
  Propagation: Application to Orbital Mechanics,''  {\em AIAA Guidance,
  Navigation, and Control Conference}, No.~2009-6082, Chicago, Illinois, 10-13
  August 2009, 10.2514/6.2009-6082.

\bibitem{terejanu2008uncertainty}
G.~Terejanu, P.~Singla, T.~Singh, and P.~D. Scott, ``Uncertainty Propagation
  for Nonlinear Dynamic Systems Using Gaussian Mixture Models,''  {\em Journal
  of Guidance, Control, and Dynamics}, Vol.~31, No.~6, 2008, pp.~1623--1633,
  10.2514/1.36247.

\bibitem{Julier1997}
S.~J. Julier and J.~K. Uhlmann, ``New extension of the Kalman filter to
  nonlinear systems,''  {\em Signal processing, sensor fusion, and target
  recognition VI}, Vol.~3068, International Society for Optics and Photonics,
  1997, pp.~182--193, 10.1117/12.280797.

\bibitem{Jones2013}
B.~A. Jones, A.~Doostan, and G.~H. Born, ``Nonlinear Propagation of Orbit
  Uncertainty Using Non-Intrusive Polynomial Chaos,''  {\em Journal of
  Guidance, Control, and Dynamics}, Vol.~36, No.~2, 2013, pp.~430--444,
  10.2514/1.57599.

\bibitem{Milani2005}
A.~Milani, M.~E. Sansaturio, G.~Tommei, O.~Arratia, and S.~R. Chesley,
  ``Multiple solutions for asteroid orbits: computational procedure and
  applications,''  {\em Astronomy \& Astrophysics}, Vol.~431, No.~2, 2005,
  pp.~729--746, 10.1051/0004-6361:20041737.

\bibitem{Tardioli2015}
C.~Tardioli, M.~Kubicek, M.~Vasile, E.~Minisci, and A.~Riccardi, ``Comparison
  of non-intrusive approaches to uncertainty propagation in orbital
  mechanics,''  {\em AAS/AIAA Astrodynamics Specialist Conference 2015},
  No.~AAS 15-545, Vail, California, American Astronautical Society, 2015,
  pp.~3979--3992.

\bibitem{vittaldev2016spacecraft}
V.~Vittaldev, R.~P. Russell, and R.~Linares, ``Spacecraft Uncertainty
  Propagation Using Gaussian Mixture Models and Polynomial Chaos Expansions,''
  {\em Journal of Guidance, Control, and Dynamics}, Vol.~39, No.~12, 2016,
  pp.~2615--2626, 10.2514/1.G001571.

\bibitem{junkins1996non}
J.~L. Junkins, M.~R. Akella, and K.~T. Alfriend, ``Non-Gaussian error
  propagation in orbital mechanics,''  {\em The Journal Of the Astronautical
  Sciences}, Vol.~44(4), 1996, pp.~541--563.

\bibitem{Sabol2010}
C.~Sabol, T.~Sukut, K.~Hill, K.~T. Alfriend, B.~Wright, Y.~Li, and
  P.~Schumacher, ``Linearized orbit covariance generation and propagation
  analysis via simple Monte Carlo simulations,''  {\em {AAS 10-134, AAS/AIAA
  Space Flight Mechanics Conference}}, San Diego, CA, February, 2010,
  pp.~14--17.

\bibitem{Folcik2011}
Z.~Folcik, A.~Lue, and J.~Vatsky, ``Reconciling covariances with reliable
  orbital uncertainty,''  {\em 12th Advanced Maui Optical and Space
  Surveillance Technologies Conference}, 2011.

\bibitem{HernandoAyuso2017}
J.~Hernando-Ayuso and C.~Bombardelli, ``Orbit covariance propagation via
  quadratic-order state transition matrix in curvilinear coordinates,''  {\em
  Celestial Mechanics and Dynamical Astronomy}, Vol.~129, Sep 2017,
  pp.~215--234, 10.1007/s10569-017-9773-9.

\bibitem{Coppola2015}
V.~T. Coppola and S.~Tanygin, ``Using Bent Ellipsoids to Represent Large
  Position Covariance in Orbit Propagation,''  {\em Journal of Guidance,
  Control, and Dynamics}, Vol.~38, No.~9, 2015, pp.~1775--1784,
  10.2514/1.G001011.

\bibitem{Melton2000}
R.~G. Melton, ``Time-Explicit Representation of Relative Motion Between
  Elliptical Orbits,''  {\em Journal of Guidance, Control, and Dynamics},
  Vol.~23, No.~4, 2000, pp.~604--610, 10.2514/2.4605.

\bibitem{Lane2006}
C.~M. Lane and P.~Axelrad, ``Formation Design in Eccentric Orbits Using
  Linearized Equations of Relative Motion,''  {\em Journal of Guidance,
  Control, and Dynamics}, Vol.~29, No.~1, 2006, pp.~146--160, 0.2514/1.13173.

\bibitem{Hill2008}
K.~Hill, K.~Alfriend, and C.~Sabol, ``Covariance-based uncorrelated track
  association,''  {\em AIAA/AAS Astrodynamics Specialist Conference}, No.~AIAA
  2008-7211, Honolulu, Hawaii, 18-21 August 2008.

\bibitem{Weis2018}
L.~M. Weis, ``Uncertainty in KS Space with Arbitrary Forces,''  {\em AIAA
  SciTech Forum}, American Institute of Aeronautics and Astronautics, Jan.
  2018, 10.2514/6.2018-0728.

\bibitem{Roa2015OD}
J.~Roa~Vicens and J.~Pel{\'a}ez~{\'A}lvarez, ``Efficient trajectory propagation
  for orbit determination problems,''  {\em AAS/AIAA Astrodynamics Specialist
  Conference}, No.~AAS-15-730, Vail, Colorado, USA, Univelt, August 2015.

\bibitem{AHA_2021}
J.~M. Aristoff, J.~T. Horwood, and K.~T. Alfriend, ``On a set of {$J_2$}
  equinoctial orbital elements and their use for uncertainty propagation,''
  {\em Celestial Mechanics and Dynamical Astronomy}, Vol.~133, No.~9, 2021,
  pp.~1--19, 10.1007/s10569-021-10004-0.

\bibitem{bau2021generalization}
G.~Ba{\`u}, J.~Hernando-Ayuso, and C.~Bombardelli, ``A generalization of the
  equinoctial orbital elements,''  {\em Celestial Mechanics and Dynamical
  Astronomy}, Vol.~133, No.~50, 2021, 10.1007/s10569-021-10049-1.

\bibitem{Arsenault_1970}
J.~L. Arsenault, K.~C. Ford, and P.~E. Koskela, ``Orbit {D}etermination {U}sing
  {A}nalytic {P}artial {D}erivatives of {P}erturbed {M}otion,''  {\em AIAA
  Journal}, Vol.~8, No.~1, 1970, pp.~4--9, 10.2514/3.5597.

\bibitem{BC_1972}
R.~A. Broucke and P.~J. Cefola, ``On the equinoctial orbit elements,''  {\em
  Celestial Mechanics}, Vol.~5, 1972, pp.~303--310, 10.1007/BF01228432.

\bibitem{Horwood2011}
J.~T. Horwood, N.~D. Aragon, and A.~B. Poore, ``Gaussian Sum Filters for Space
  Surveillance: Theory and Simulations,''  {\em Journal of Guidance, Control,
  and Dynamics}, Vol.~34, November 2011, pp.~1839--1851, 10.2514/1.53793.

\bibitem{battin1999introduction}
R.~Battin, {\em {An Introduction to the Mathematics and Methods of
  Astrodynamics}}.
\newblock Reston, Virginia: AIAA, 1999.

\bibitem{brouwer1959solution}
D.~Brouwer, ``Solution of the problem of artificial satellite theory without
  drag,''  {\em The Astronomical Journal}, Vol.~64, 1959, p.~378,
  10.1086/107958.

\bibitem{abbasrezaee2019conceptual}
P.~Abbasrezaee, M.~Mirshams, and S.~Seyed-Zamani, ``Conceptual GEO
  Telecommunication All-Electric Satellite Design Based on Statistical Model,''
   {\em 2019 9th International Conference on Recent Advances in Space
  Technologies (RAST)}, IEEE, 2019, pp.~503--507, 10.1109/RAST.2019.8767854.

\bibitem{ries2016development}
J.~Ries, S.~Bettadpur, R.~Eanes, Z.~Kang, U.~Ko, C.~McCullough, P.~Nagel,
  N.~Pie, S.~Poole, T.~Richter, {\em et~al.}, ``Development and Evaluation of
  the Global Gravity Model GGM05-CSR-16-02,''  {\em Center for Space Research,
  Univ. of Texas at Austin TR CSR-16-02, Austin, TX}, 2016.

\bibitem{cunningham1970computation}
L.~E. Cunningham, ``On the computation of the spherical harmonic terms needed
  during the numerical integration of the orbital motion of an artificial
  satellite,''  {\em Celestial Mechanics}, Vol.~2, No.~2, 1970, pp.~207--216,
  10.1007/BF01229495.

\bibitem{metris1998derivatives}
G.~M{\'e}tris, J.~Xu, and I.~Wytrzyszczak, ``Derivatives of the gravity
  potential with respect to rectangular coordinates,''  {\em Celestial
  Mechanics and Dynamical Astronomy}, Vol.~71, No.~2, 1998, pp.~137--151,
  10.1023/A:1008361202235.

\bibitem{fantino2009methods}
E.~Fantino and S.~Casotto, ``Methods of harmonic synthesis for global
  geopotential models and their first-, second-and third-order gradients,''
  {\em Journal of Geodesy}, Vol.~83, No.~7, 2009, pp.~595--619,
  10.1007/s00190-008-0275-0.

\end{thebibliography}

\appendix

\section{Computation of $\partial{\bf f}/\partial{\bf y}$}

\label{sec:dfdy}

In the following, the elements of the $6\times6$ matrix $\partial{\bf
  f}/\partial{\bf y}$ are computed, where ${\bf f}({\bf y},t)$ and
${\bf y}$ have been introduced in Section \ref{sec:LCP}. For this
purpose, it is useful to define the function $\bm{\mathfrak{f}}({\bf
  y},\,{\bf a},\,{\bf b},\,t)$:
\[
\bm{\mathfrak{f}}({\bf y},\,{\bf a}({\bf y},t),\,{\bf b}({\bf y},t),\,t)={\bf f}({\bf y},t),
\]
where 
\begin{align*}
  {\bf a} & =(r,\,u,\,h,\,c,\,L)^{T},\\
  {\bf b} & =(\mathscr{U}_{t},\,P_{r},\,P_{f},\,F_{h},\,S)^{T},
\end{align*}
with 
\[
  {U}_{t}=\frac{\partial{\mathscr{U}}}{\partial t},\qquad
  S=2\mathscr{U}-rF_{r}.
\]
Recall that $L$ is the true longitude (Eq.~\ref{eq:tr_long}) and
$F_{r}$, $F_{h}$, $P_{r}$, $P_{f}$ are defined in
Eq.~\eqref{eq:FPprojs}.

The components of $\bm{\mathfrak{f}}$ are given by equations
\eqref{eq:nudot}--\eqref{eq:q2dot}.  One has
\begin{equation}
  \frac{\partial{\bf f}}{\partial{\bf y}}=\frac{\partial\bm{\mathfrak{f}}}{\partial{\bf y}}+
  \frac{\partial\bm{\mathfrak{f}}}{\partial{\bf a}}\frac{\partial{\bf a}}{\partial{\bf y}}+
  \frac{\partial\bm{\mathfrak{f}}}{\partial{\bf b}}\frac{\partial{\bf b}}{\partial{\bf y}}.
  \label{eq:1}
\end{equation}
The following subsections deal with the computation of the matrices
that appear on the right-hand side of (\ref{eq:1}).

It is useful to define the quantities 
\begin{align*}
  \sigma_{1} & =p_{2}+\cos L, & \sigma_{2} & =p_{1}+\sin L,\\
  \sigma_{3} & =\sin L+\varsigma\sigma_{1}, & \sigma_{4} & =\cos L+\varsigma\sigma_{2},\\
  \varsigma & =\frac{r}{\varrho}, & \tilde{\varsigma} & =1+\frac{r}{\varrho},
\end{align*}
where $\varrho$ is defined in \eqref{eq:a_nu}.

\subsection{$\partial\bm{\mathfrak{f}}/\partial{\bf y}$}

The six rows of this matrix are given by\footnote{Here $Y_{i}$ denotes
  the $i$-th component of the vector ${\bf Y}$.}:
\begin{align*}
  \frac{\partial\mathfrak{f}_{1}}{\partial{\bf y}} & =
  -\left(\frac{1}{\mu\nu}\right)^{2/3}\dot{\mathscr{E}}(1,\,\,0\,\,,0\,\,,0\,\,,0\,\,,0),\\[1ex]
  \frac{\partial\mathfrak{f}_{2}}{\partial{\bf y}} & =
  \Bigl(\frac{2X}{3\mu\beta}S,\,\,\frac{r^{2}}{c^{2}}\dot{\mathscr{E}},\,\,\frac{h-c}{r^{2}}-
  \dfrac{\gamma}{h}F_{h}+\frac{2}{c}S,\,\,0,\,\,-\frac{Xp_{2}}{h}F_{h},\,\,\frac{Yp_{2}}{h}F_{h}\Bigr),\\
  \frac{\partial\mathfrak{f}_{3}}{\partial{\bf y}} & =
  \Bigl(-\frac{2Y}{3\mu\beta}S,\,\,\dfrac{\gamma}{h}F_{h}-\frac{h-c}{r^{2}}-\frac{2}{c}S,\,\,
  \frac{r^{2}}{c^{2}}\dot{\mathscr{E}},\,\,0,\,\,\frac{Xp_{1}}{h}F_{h},\,\,-\frac{Yp_{1}}{h}F_{h}\Bigr),\\
  \frac{\partial\mathfrak{f}_{4}}{\partial{\bf y}} & =
  \Bigl(1-\frac{2\alpha r}{3\mu\beta}S,\,\,\frac{\partial\mathfrak{f}_{4}}{\partial p_{1}},\,\,
  \frac{\partial\mathfrak{f}_{4}}{\partial p_{2}},\,\,0,-\frac{X}{h}F_{h},\,\,\frac{Y}{h}F_{h}\Bigr),\\
  \frac{\partial\mathfrak{f}_{5}}{\partial{\bf y}} & =
  \Bigl(0,\,\,0,\,\,0,\,\,0,\,\,\frac{Yq_{1}}{h}F_{h},\,\,\frac{Yq_{2}}{h}F_{h}\Bigr),\\[1ex]
  \frac{\partial\mathfrak{f}_{6}}{\partial{\bf y}} & =
  \Bigl(0,\,\,0,\,\,0,\,\,0,\,\,\frac{Xq_{1}}{h}F_{h},\,\,\frac{Xq_{2}}{h}F_{h}\Bigr),
\end{align*}
where 
\[
\frac{\partial\mathfrak{f}_{4}}{\partial p_{i}}=
\biggl[\frac{ruc}{\mu^{2}}\tilde{\varsigma}\dot{\mathscr{E}}+\Bigl(1-\frac{r}{\textsl{\textrm{a}}}-
  \frac{1}{\alpha^{2}}\Bigr)\frac{S}{c}\biggr]\frac{\alpha^{2}p_{i}}{\beta},\quad i=1,2.
\]
The definitions of $X$, $Y$ are given in~\eqref{eq:XY} and those of
$\gamma$, $\alpha$, $\beta$ in~\eqref{eq:g_al_be}.

\subsection{$\partial\bm{\mathfrak{f}}/\partial{\bf a}$}

The six rows of this matrix are given by:
\begin{enumerate}
\item[1$^{st}$] row 
  \[
  \frac{\partial\mathfrak{f}_{1}}{\partial{\bf a}}=-3\Bigl(\frac{\nu}{\mu^{2}}\Bigr)^{1/3}
  \Bigl(-\frac{h}{r^{2}}P_{f},\,\,P_{r},\,\,\frac{1}{r}P_{f},\,\,0,\,\,0\Bigr);
  \]
\item[2$^{nd}$] row 
  \begin{align*}
    \frac{\partial\mathfrak{f}_{2}}{\partial a_{1}} & =
    -\biggl[\frac{2(h-c)}{r^{2}}+\frac{\gamma}{h}F_{h}\biggr]\frac{p_{2}}{r}+\frac{\cos L}{c\textsl
    {\textrm{a}}}S+\frac{1}{\mu}[\sigma_{3}(\mathscr{U}_{t}+uP_{r})+\varsigma\sigma_{2}\dot{\mathscr{E}}],\\
    \frac{\partial\mathfrak{f}_{2}}{\partial a_{2}} & = \frac{r\sigma_{3}}{\mu}P_{r},\\[1ex]
    \frac{\partial\mathfrak{f}_{2}}{\partial a_{3}} & = \biggl(\frac{1}{r^{2}}+\frac{\gamma}
    {h^{2}}F_{h}\biggr)p_{2}+\frac{\sigma_{3}}{\mu}P_{f},\\[1ex]
    \frac{\partial\mathfrak{f}_{2}}{\partial a_{4}} & =
    -\frac{p_{2}}{r^{2}}-\frac{1}{c^{2}}\biggl[\biggl(\frac{X}{\textsl{\textrm{a}}}+2p_{2}\biggr)S+
    \frac{2r^{2}\sigma_{2}}{c}\dot{\mathscr{E}}\biggr],\\
    \frac{\partial\mathfrak{f}_{2}}{\partial a_{5}} & =
    (q_{1}\sin L+q_{2}\cos L)\frac{rp_{2}}{h}F_{h}-\frac{Y}{c\textsl{\textrm{a}}}S+
    \frac{X\tilde{\varsigma}}{\mu}\dot{\mathscr{E}};
  \end{align*}
\item[3$^{rd}$] row 
  \begin{align*}
    \frac{\partial\mathfrak{f}_{3}}{\partial a_{1}} & =
    \biggl[\frac{2(h-c)}{r^{2}}+\frac{\gamma}{h}F_{h}\biggr]\frac{p_{1}}{r}-\frac{\sin L}{c\textsl
    {\textrm{a}}}S+\frac{1}{\mu}[\sigma_{4}(\mathscr{U}_{t}+uP_{r})+\varsigma\sigma_{1}\dot{\mathscr{E}}],\\
    \frac{\partial\mathfrak{f}_{3}}{\partial a_{2}} & = \frac{r\sigma_{4}}{\mu}P_{r},\\
    \frac{\partial\mathfrak{f}_{3}}{\partial a_{3}} & = -\biggl(\frac{1}{r^{2}}+\frac{\gamma}
    {h^{2}}F_{h}\biggr)p_{1}+\frac{\sigma_{4}}{\mu}P_{f},\\[1ex]
    \frac{\partial\mathfrak{f}_{3}}{\partial a_{4}} & =
    \frac{p_{1}}{r^{2}}+\frac{1}{c^{2}}\biggl[\biggl(\frac{Y}{\textsl{\textrm{a}}}+2p_{1}\biggr)S
    -\frac{2r^{2}\sigma_{1}}{c}\dot{\mathscr{E}}\biggr],\\
    \frac{\partial\mathfrak{f}_{3}}{\partial a_{5}} & =
    -(q_{1}\sin L+q_{2}\cos L)\frac{rp_{1}}{h}F_{h}-\frac{X}{c\textsl{\textrm{a}}}S
    -\frac{Y\tilde{\varsigma}}{\mu}\dot{\mathscr{E}};
  \end{align*}
\item[4$^{th}$] row 
  \begin{align*}
    \frac{\partial\mathfrak{f}_{4}}{\partial a_{1}} & =
    -\frac{2(h-c)}{r^{3}}-\frac{\gamma}{rh}F_{h}+\frac{uc\alpha}{\mu^{2}}\left[\tilde{\varsigma}(\mathscr{U}_{t}+
    uP_{r})+\varsigma\dot{\mathcal{E}}\right]-\frac{\alpha}{c\textsl{\textrm{a}}}S,\\[1ex]
    \frac{\partial\mathfrak{f}_{4}}{\partial a_{2}} & =
    \frac{rc\alpha}{\mu^{2}}\tilde{\varsigma}(\dot{\mathscr{E}}+uP_{r}),\\[1ex]
    \frac{\partial\mathfrak{f}_{4}}{\partial a_{3}} & =
    \frac{1}{r^{2}}+\frac{\gamma}{h^{2}}F_{h}+\frac{uc\alpha}{\mu^{2}}\tilde{\varsigma}P_{f},\\[1ex]
    \frac{\partial\mathfrak{f}_{4}}{\partial a_{4}} & =
    -\frac{1}{r^{2}}+\frac{ru\alpha}{\mu^{2}}(1-\varsigma)\dot{\mathscr{E}}-\biggl[\frac{1}{\alpha}+
    \alpha\Bigl(1-\frac{r}{\textsl{\textrm{a}}}\Bigr)\biggr]\frac{S}{c^{2}},\\[1ex]
    \frac{\partial\mathfrak{f}_{4}}{\partial a_{5}} & =\frac{r}{h}(q_{1}\sin L+q_{2}\cos L)F_{h};
  \end{align*}
\item[5$^{th}$] row 
  \[
  \frac{\partial\mathfrak{f}_{5}}{\partial{\bf a}}=
  \frac{F_{h}}{2h}(1+q_{1}^{2}+q_{2}^{2})\Bigl(\sin L,\,\,0,\,\,-\frac{Y}{h},\,\,0,\,\,X\Bigr);
  \]
\item[6$^{th}$] row 
  \[
  \frac{\partial\mathfrak{f}_{6}}{\partial{\bf a}}=
  \frac{F_{h}}{2h}(1+q_{1}^{2}+q_{2}^{2})\Bigl(\cos L,\,\,0,\,\,-\frac{X}{h},\,\,0,\,\,-Y\Bigr).
  \]
\end{enumerate}

\subsection{$\partial{\bf a}/\partial{\bf y}$}

The five rows of this matrix are given by: 
\begin{enumerate}
\item[1$^{st}$] row 
  \[
  \frac{\partial a_{1}}{\partial{\bf y}} =
  \Bigl(-\frac{2r}{3\nu},\,\,-\frac{u}{\nu}\cos\mathcal{K}-\textsl{\textrm{a}}\sin\mathcal{K},\,\,
  \frac{u}{\nu}\sin\mathcal{K}-\textsl{\textrm{a}}\cos\mathcal{K},\,\,\frac{u}{\nu},\,\,0,\,\,0\Bigr);
  \]
\item[2$^{nd}$] row 
  \begin{align*}
    \frac{\partial a_{2}}{\partial{\bf y}} & =
    \biggl(\frac{u}{3\nu},\,\,\frac{1}{r\nu}\Bigl(u^{2}-\frac{\mu}{r}\Bigr)\cos\mathcal{K}+
    \frac{u\textsl{\textrm{a}}}{r}\sin\mathcal{K},\,\,\frac{1}{r\nu}\Bigl(\frac{\mu}{r}-u^{2}\Bigr)
    \sin\mathcal{K}\\[0.5ex]
    & \quad\,+\frac{u\textsl{\textrm{a}}}{r}\cos\mathcal{K},\,\,\frac{\sqrt{\mu\textsl{\textrm{a}}}}{r}
    \Bigl(\frac{\textsl{\textrm{a}}}{r}-1\Bigr)-\frac{u^{2}}{r\nu},\,\,0,\,\,0\biggr);
  \end{align*}
\item[3$^{rd}$] row 
  \[
  \frac{\partial a_{3}}{\partial{\bf y}} =
  \frac{1}{h}\biggl(c\frac{\partial a_{4}}{\partial{\bf y}}-2r\mathscr{U}\frac{\partial a_{1}}{\partial{\bf y}}-
  r^{2}\frac{\partial\mathscr{U}}{\partial{\bf y}}\biggr);
  \]
\item[4$^{th}$] row 
  \[
  \frac{\partial a_{4}}{\partial{\bf y}} =
  \biggl(-\frac{\textsl{\textrm{a}}^{2}\beta}{3},\,\,-\Bigl(\frac{\mu^{2}}{\nu}\Bigr)^{1/3}\frac{p_{1}}{\beta},
  \,\,-\Bigl(\frac{\mu^{2}}{\nu}\Bigr)^{1/3}\frac{p_{2}}{\beta},\,\,0,\,\,0,\,\,0\biggr);
  \]
\item[5$^{th}$] row 
  \begin{align*}
    \frac{\partial a_{5}}{\partial\nu} & =
    \frac{\partial a_{5}}{\partial q_{1}} = \frac{\partial a_{5}}{\partial q_{2}}=0,\\[1ex]
    \frac{\partial a_{5}}{\partial p_{1}} & =
    \frac{\textsl{\textrm{a}}}{r}\Bigl[\alpha\Bigl(\frac{\textsl{\textrm{a}}}{r}\cos\mathcal{K}-
    \frac{ru\alpha p_{1}}{c}\Bigr)\Bigl(\frac{1}{\varsigma}-1\Bigr)-\frac{Yu\alpha}
    {\sqrt{\mu\textsl{\textrm{a}}}}\\[0.5ex]
    & \quad\,-\cos L-\frac{\textsl{\textrm{a}}}{r}\cos\mathcal{K}\cos(L-\mathcal{K})\Bigr],\\[1ex]
    \frac{\partial a_{5}}{\partial p_{2}} & =
    \frac{\textsl{\textrm{a}}}{r}\Bigl[-\alpha\Bigl(\frac{\textsl{\textrm{a}}}{r}\sin\mathcal{K}+
    \frac{ru\alpha p_{2}}{c}\Bigr)\Bigl(\frac{1}{\varsigma}-1\Bigr)-\frac{Xu\alpha}
    {\sqrt{\mu\textsl{\textrm{a}}}}\\[0.5ex]
    & \quad\,+\sin L+\frac{\textsl{\textrm{a}}}{r}\sin\mathcal{K}\cos(L-\mathcal{K})\Bigr],\\[1ex]
    \frac{\partial a_{5}}{\partial\mathcal{L}} & =
    \frac{\textsl{\textrm{a}}^{2}}{r^{2}}\Bigl[\alpha\Bigl(\frac{r}{\textsl{\textrm{a}}}-1\Bigr)
    \Bigl(\frac{1}{\varsigma}-1\Bigr)+\cos(L-\mathcal{K})\Bigr].
  \end{align*}
\end{enumerate}
The quantity $\mathcal{K}$ can be computed as described in
\cite{bau2021generalization}, Sect. 4. For the computation of
$\partial\mathscr{U}/\partial{\bf y}$ see Sect.~\ref{sec:dbdy}.

\subsection{$\partial\bm{\mathfrak{f}}/\partial{\bf b}$}

The six rows of this matrix are given by: 
\begin{enumerate}
\item[1$^{st}$] row 
\[
\frac{\partial\mathfrak{f}_{1}}{\partial{\bf b}} =
-3\Bigl(\frac{\nu}{\mu^{2}}\Bigr)^{1/3}\Bigl(1,\,\,u,\,\,\frac{h}{r},\,\,0,\,\,0\Bigr);
\]
\item[2$^{nd}$] row 
\begin{align*}
  \frac{\partial\mathfrak{f}_{2}}{\partial{\bf b}} & =
  \biggl(\frac{r}{\mu}\sigma_{3},\,\,\frac{ru}{\mu}\sigma_{3},\,\,\frac{h}{\mu}\sigma_{3},\,\,
  -\frac{\gamma p_{2}}{h},\,\,\frac{1}{c}\biggl(\frac{X}{\textsl{\textrm{a}}}+2p_{2}\biggr)\biggr);
\end{align*}
\item[3$^{rd}$] row 
\begin{align*}
  \frac{\partial\mathfrak{f}_{3}}{\partial{\bf b}} & =
  \biggl(\frac{r}{\mu}\sigma_{4},\,\,\frac{ru}{\mu}\sigma_{4},\,\,\frac{h}{\mu}\sigma_{4},\,\,
  \frac{\gamma p_{1}}{h},\,\,-\frac{1}{c}\biggl(\frac{Y}{\textsl{\textrm{a}}}+2p_{1}\biggr)\biggr);
\end{align*}
\item[4$^{th}$] row 
\begin{align*}
  \frac{\partial\mathfrak{f}_{4}}{\partial{\bf b}} & =
  \biggl(\frac{ruc\alpha}{\mu^{2}}\tilde{\varsigma},\,\,\frac{ru^{2}c\alpha}{\mu^{2}}\tilde{\varsigma},\,\,
  \frac{uhc\alpha}{\mu^{2}}\tilde{\varsigma},\,\,-\frac{\gamma}{h},\,\,\frac{1}{c}\biggl[\frac{1}{\alpha}+
  \alpha\Bigl(1-\frac{r}{\textsl{\textrm{a}}}\Bigr)\biggr]\biggr);
\end{align*}
\item[5$^{th}$] row 
\[
\frac{\partial\mathfrak{f}_{5}}{\partial{\bf b}} =
\biggl(0,\,\,0,\,\,0,\,\,\frac{Y}{2h}(1+q_{1}^{2}+q_{2}^{2}),\,\,0\biggr);
\]
\item[6$^{th}$] row 
\[
\frac{\partial\mathfrak{f}_{6}}{\partial{\bf b}} =
\biggl(0,\,\,0,\,\,0,\,\,\frac{X}{2h}(1+q_{1}^{2}+q_{2}^{2}),\,\,0\biggr).
\]
\end{enumerate}

\subsection{$\partial{\bf b}/\partial{\bf y}$\label{subsec:dbdy}}

\label{sec:dbdy} Let 
\[
{\bf x}=(x,\,y,\,z,\,\dot{x},\,\dot{y},\,\dot{z})^{T}
\]
be the set of Cartesian coordinates of the position ${\bf r}$ and
velocity $\dot{{\bf r}}$ of the propagated body with respect to an
inertial reference frame $\Sigma$. In order to compute $\partial{\bf
  b}/\partial{\bf y}$ the chain rule is applied:
\[
\frac{\partial{\bf b}}{\partial{\bf y}} =
\frac{\partial{\bf b}}{\partial{\bf x}}\frac{\partial{\bf x}}{\partial{\bf y}}.
\]
Note that the matrix $\partial{\bf x}/\partial{\bf y}$ is provided in
\cite{bau2021generalization}, Sect. 6.1. The first row of the matrix
$\partial{\bf b}/\partial{\bf x}$ is
\[
\frac{\partial b_{1}}{\partial{\bf x}} =
\left(\frac{\partial\mathscr{U}_{t}}{\partial{\bf r}},\,0,\,0,\,0\right).
\]
Regarding the other rows one has: 
\begin{enumerate}
\item[2$^{nd}$] row 
  \[
  \frac{\partial b_{2}}{\partial{\bf x}} =
  {\bf e}_{r}^{T}\frac{\partial{\bf P}}{\partial{\bf x}}+{\bf P}^{T}\frac{\partial{\bf e}_{r}}{\partial{\bf x}},
  \]
\item[3$^{rd}$] row 
  \[
  \frac{\partial b_{3}}{\partial{\bf x}} =
  {\bf e}_{f}^{T}\frac{\partial{\bf P}}{\partial{\bf x}}+{\bf P}^{T}\frac{\partial{\bf e}_{f}}{\partial{\bf x}},
  \]
\item[4$^{th}$] row 
  \[
  \frac{\partial b_{4}}{\partial{\bf x}} =
  {\bf e}_{h}^{T}\frac{\partial{\bf F}}{\partial{\bf x}}+{\bf F}^{T}\frac{\partial{\bf e}_{h}}{\partial{\bf x}},
  \]
\item[5$^{th}$] row 
  \[
  \frac{\partial b_{5}}{\partial{\bf x}} =
  2\frac{\partial\mathscr{U}}{\partial{\bf x}}-\frac{\partial r}{\partial{\bf x}}F_{r}-r\Bigl({\bf e}_{r}^{T}
  \frac{\partial{\bf F}}{\partial{\bf x}}+{\bf F}^{T}\frac{\partial{\bf e}_{r}}{\partial{\bf x}}\Bigr),
  \]
\end{enumerate}
where the two vectors ${\bf F}$ and ${\bf P}$ (introduced in
Eqs.~\ref{eq:eqm}, \ref{eq:FP}) are expressed in $\Sigma$ and
\[
\frac{\partial r}{\partial{\bf x}}=({\bf e}_{r}^{T},\,0,\,0,\,0),
\]
\begin{align*}
  \frac{\partial{\bf e}_{r}}{\partial{\bf r}} & =
  -\frac{1}{r^{3}}R^{2}, & \frac{\partial{\bf e}_{r}}{\partial\dot{{\bf r}}} & =O_{3},\\[1ex]
  \frac{\partial{\bf e}_{f}}{\partial{\bf r}} & =
  \frac{1}{rh}\left(\frac{u}{r}R^{2}+\frac{1}{h^{2}}HCH^{T}\right), & \frac{\partial{\bf e}_{f}}
  {\partial\dot{{\bf r}}} & =-\frac{1}{rh}\left(R^{2}+\frac{1}{h^{2}}HBH^{T}\right),\\[1ex]
  \frac{\partial{\bf e}_{h}}{\partial{\bf r}} & =
  \frac{1}{h^{2}}\left(H\dot{{\bf r}}\right){\bf e}_{h}^{T}, &
  \frac{\partial{\bf e}_{h}}{\partial\dot{{\bf r}}} & =-\frac{r}{h}{\bf e}_{f}{\bf e}_{h}^{T},
\end{align*}
with $O_{3}$, $I_{3}$ denoting the null and identity $3\times3$
matrices, respectively, and
\[
R=\left(
\begin{array}{ccc}
  \phantom{-}0 & -z & \phantom{-}y\\
  \phantom{-}z & \phantom{-}0 & -x\\
  -y & \phantom{-}x & \phantom{-}0
\end{array}
\right),\qquad
H=\left(
\begin{array}{ccc}
  \phantom{-}0 & -h_{3} & \phantom{-}h_{2}\\
  \phantom{-}h_{3} & \phantom{-}0 & -h_{1}\\
  -h_{2} & \phantom{-}h_{1} & \phantom{-}0
\end{array}
\right),
\]
\[
C=\dot{{\bf r}}{\bf r}^{T},\qquad B={\bf r}{\bf r}^{T},
\]
\[
h_{1}=y\dot{z}-z\dot{y},\qquad h_{2}=z\dot{x}-x\dot{z},\qquad h_{3}=x\dot{y}-y\dot{x}.
\]
Moreover, for a generic vector ${\bf Y}$ one has 
\[
\frac{\partial{\bf Y}}{\partial{\bf x}} =
\left(\frac{\partial{\bf Y}}{\partial{\bf r}}\,\vert\,\frac{\partial{\bf Y}}{\partial\dot{{\bf r}}}\right).
\]
Finally, assuming that $\mathscr{U}$ is a function of ${\bf r}$ and
possibly of time, one has
\[
\frac{\partial\mathscr{U}}{\partial{\bf y}} =
\frac{\partial\mathscr{U}}{\partial{\bf r}}\frac{\partial{\bf r}}{\partial{\bf y}},
\]
where $\partial{\bf r}/\partial{\bf y}$ is given in
\cite{bau2021generalization}, Sect. 6.1.

\section{Secular evolution of $\mathcal{L}$ under the effect of J$_{2}$}
\label{sec:secL_J2}

From Eqs.~\eqref{eq:dL_J2},~\eqref{eq:hmc} one can write, for an
equatorial orbit
\begin{equation}
  \dot{\mathcal{L}} = \nu_{0}+\frac{J_{2}\mu R^{2}}{2cr^{3}}\biggl[1+\frac{1}{\alpha}+
  \alpha\biggl(1-\frac{r}{\textrm{a}_{0}}\biggr)\biggr],
  \label{eq:dL_J2_2}
\end{equation}
where terms of order higher than one in J$_2$ have been neglected in
Eq.~(\ref{eq:hmc}). Let the mean value of the eccentricity $e$ be
equal to zero. Then, from Eq.~\eqref{eq:ecc_gen} one can assume that
the mean generalized eccentricity is also zero, which implies
$\alpha=1/2$, $c=\sqrt{\mu\textrm{a}_{0}}$, $r=\textrm{a}_{0}$. After
applying these substitutions, Eq.~\eqref{eq:dL_J2_2} reduces to
\[
\dot{\mathcal{L}}=\nu_{0}+\frac{3J_{2}\mu R^{2}}{2\textrm{a}_{0}^{3}\sqrt{\mu\textrm{a}_{0}}},
\]
where the symbol $\dot{\mathcal{L}}$ denotes now the secular rate of
the generalized mean longitude. Using the first relation
in~\eqref{eq:a_nu} and the definition of $A$ in~\eqref{eq:A} one finds
\[
\dot{\mathcal{L}}=\nu_{0}+\frac{3J_{2}R^{2}\nu_{0}^{7/3}}{2\mu^{2/3}}=\nu_{0}\bigl(1+J_{2}A\nu_{0}^{4/3}\bigr),
\]
which can be integrated to give Eq.~\eqref{eq:elleg_t}.

\section{Partial derivatives of the spherical harmonics potential}
\label{sec:shp_der}

Multiple algorithms exist to compute the potential of the spherical
harmonics perturbation and its derivatives. A non-exhaustive list
includes the methods of Legendre, Clenshaw, Pines and the one used in
this work, Cunningham-Metris. A comparison of these methods is
presented in \cite{fantino2009methods}. All of them are naturally
defined in ECEF coordinates, and can provide the potential and its
first and second derivatives with respect to the ECEF Cartesian
position. In order to propagate the orbital state and the state
transition matrix in the GEqOE space one has to compute
\begin{equation}
  \frac{\partial\mathscr{U}}{\partial\boldsymbol{r}_{\text{ECI}}},\quad
  \frac{\partial\mathscr{U}}{\partial t},\quad \frac{\partial^{2}\mathscr{U}}
  {\partial\boldsymbol{r}_{\text{ECI}}^{2}},\quad
  \frac{\partial}{\partial\boldsymbol{r}_{\text{ECI}}}\biggl(\frac{\partial\mathscr{U}}{\partial t}\biggr),
  \label{eq:terms}
\end{equation}
where ${\bf r}_{\text{ECI}}$ denotes the position of the propagated
body expressed in the Earth-centered inertial frame (ECI). The first
two terms in \eqref{eq:terms} appear in the equations of motion, while
the other two are needed for the propagation of the state transition
matrix.

The Cartesian state vector ${\bf x}$ can be rotated from ECI to ECEF
as
\[
  {\bf x}_{\text{ECEF}}=\left[
  \begin{array}{cc}
    \mathfrak{R}(t) & O_3\\
    \dot{\mathfrak{R}}(t) & \mathfrak{R}(t)
  \end{array}
  \right]{\bf x}_{\text{ECI}},
  \label{eq:ECI_to_ECEF}
\]
where $\mathfrak{R}$ is the rotation matrix from ECI to ECEF and $O_3$
is the $3\times3$ null matrix.

The first term in~\eqref{eq:terms} is used to calculate
$\boldsymbol{F}$:
\[
\frac{\partial\mathscr{U}}{\partial\boldsymbol{r}_{\text{ECI}}} =
\frac{\partial\mathscr{U}}{\partial\boldsymbol{r}_{\text{ECEF}}}\mathfrak{R}.
\]

The second term can be obtained as the time derivative of
$\mathscr{U}(t)=\mathscr{U}(\boldsymbol{r}_{\text{ECI}}(t))$ after
expressing the ECI position in ECEF coordinates:
\[
\frac{\partial\mathscr{U}}{\partial t} =
\frac{\partial\mathscr{U}}{\partial\boldsymbol{r}_{\text{ECEF}}}\dot{\mathfrak{R}}\,\mathfrak{R}^{T}
\boldsymbol{r}_{\text{ECEF}}.
\label{eq:dUdt}
\]

The third term contributes to $\partial{\bf F}/{\partial{\bf x}}$ in
Section \ref{subsec:dbdy}, and reads
\[
\frac{\partial^{2}\mathscr{U}}{\partial\boldsymbol{r}_{\text{ECI}}^{2}} =
\mathfrak{R}^{T}\frac{\partial^{2}\mathscr{U}}{\partial\boldsymbol{r}_{\text{ECEF}}^{2}}\mathfrak{R}.
\]

The last term in~\eqref{eq:terms}, which is the first component of
$\partial b_{1}/{\partial{\bf x}}$ (see Sect.~\ref{subsec:dbdy}), is
given by
\[
\frac{\partial}{\partial\boldsymbol{r}_{\text{ECI}}}
\biggl(\frac{\partial\mathscr{U}}{\partial t}\biggr) =
\mathfrak{R}^{T}\frac{\partial^{2}\mathscr{U}}{\partial\boldsymbol{r}_{\text{ECEF}}^{2}}
\dot{\mathfrak{R}}\,\mathfrak{R}^{T} \boldsymbol{r}_{\text{ECEF}}+
\frac{\partial\mathscr{U}}{\partial\boldsymbol{r}_{\text{ECEF}}}\dot{\mathfrak{R}}.
\]

\end{document}